\documentclass{emulateapj}
\usepackage{comment}

\newcommand{\lya}        {Ly$\alpha$}
\newcommand{\unitcgssb}  {erg\,s$^{-1}$\,cm$^{-2}$\,arcsec$^{-2}$}

\newcommand{\unitcgslum} {erg\,s$^{-1}$}
\newcommand{\nb}         {{\sl NB}}

\newcommand{\CIV}        {\hbox{{\rm C}\kern 0.1em{\sc iv}}}
\newcommand{\NII}        {[\ion{N}{2}]}
\newcommand{\kms}        {km\,s$^{-1}$}
\newcommand{\bptx}       {[\ion{N}{2}]/H$\alpha$}

\newcommand{\halamb}     {H$\alpha$ $\lambda$6563}
\newcommand{\dvlya}      {$\Delta v_{\rm Ly\alpha}$}
\newcommand{\dvis}       {$\Delta v_{\rm IS}$}
\newcommand{\vexp}       {$v_{\rm exp}$}
\newcommand{\NHI}        {$N_{\rm{H I}}$}
\newcommand{\msun}       {$M_{\sun}$}
\newcommand\avg[1]       {\langle#1\rangle}

\slugcomment{Accepted for publication in ApJ.}

\shorttitle{Kinematics  of Gas in  \lya\ Blobs}
\shortauthors{Yang et al.}

\begin{document}

\title{
Gas Kinematics in L\lowercase{y$\alpha$} Nebulae\altaffilmark{*}
}

\author{
Yujin Yang\altaffilmark{1},
Ann Zabludoff\altaffilmark{2},
Knud Jahnke\altaffilmark{1},
Daniel Eisenstein\altaffilmark{2,3},
Romeel Dav\'e\altaffilmark{2}, \\
Stephen A. Shectman\altaffilmark{4},
Daniel D. Kelson\altaffilmark{4}
}

%\altaffiltext{1}{Max-Planck-Institut f\"ur Astronomie, K\"onigstuhl 17, Heidelberg, Germany. yyang@mpia.de}
%\altaffiltext{2}{Steward Observatory, University of Arizona, 933 North Cherry Avenue, Tucson AZ 85721}
%\altaffiltext{3}{Department of Astronomy, Harvard University, 60 Garden Street, Cambridge, MA 02138}
%\altaffiltext{4}{Observatories of the Carnegie Institution of Washington, Pasadena, CA 91101}

\altaffiltext{1}{Max-Planck-Institut f\"ur Astronomie, K\"onigstuhl 17, Heidelberg, Germany. yyang@mpia.de}
\altaffiltext{2}{Steward Observatory, University of Arizona}
\altaffiltext{3}{Department of Astronomy, Harvard University}
\altaffiltext{4}{Observatories of the Carnegie Institution of Washington}

%\altaffiltext{1}{Max-Planck-Institut f\"ur Astronomie, Heidelberg, Germany}
%\altaffiltext{2}{Steward Observatory, University of Arizona}
%\altaffiltext{3}{Observatories of the Carnegie Institution of Washington}

\altaffiltext{*}{This paper includes data gathered with the 6.5 meter
                 Magellan Telescopes located at Las Campanas Observatory, Chile.}

\altaffiltext{*}{Based on observations made with ESO telescope at the 
                 Paranal Observatory, under the program ID 082.A-0613.}

%\email{yyang@mpia.de}

\begin{abstract}

Exploring the origin of \lya\ nebulae (``blobs") at high redshift
requires measurements of their gas kinematics that are impossible with
only the resonant, optically-thick \lya\ line.  To define gas motions
relative to the systemic velocity of the blob, the \lya\ line must be
compared with an optically-thin line like \halamb, which is not much
altered by radiative transfer effects and is more concentrated about
the galaxies embedded in the nebula's core.  We obtain optical and
near-infrared (NIR) spectra of the two brightest \lya\ blobs (CDFS-LAB01
and CDFS-LAB02) from the \citet{Yang10} sample using the Magellan/MagE
optical and VLT/SINFONI NIR spectrographs.  Both the \lya\ and H$\alpha$
lines confirm that these blobs lie at the survey redshift, $z\sim2.3$.
Within each blob, we detect several H$\alpha$ sources, which roughly
correspond to galaxies seen in {\sl HST} rest-frame UV images.
The H$\alpha$ detections show that these galaxies have large internal
velocity dispersions ($\sigma_v = 130 - 190$\,\kms) and that, in the one
system (LAB01), where we can reliably extract profiles for two H$\alpha$
sources, their velocity difference is $\Delta v$ $\sim$ 440\,\kms.
The presence of multiple galaxies within the blobs, and those galaxies'
large velocity dispersions and large relative motion, is consistent
with our previous finding that \lya\ blobs inhabit massive dark matter
halos that will evolve into those typical of rich clusters today and
that the embedded galaxies may eventually become brightest cluster
galaxies \citep{Yang10}.
To determine whether the gas near the embedded galaxies is predominantly
infalling or outflowing, we compare the \lya\ and H$\alpha$ line
centers, finding that \lya\ is not offset (\dvlya\ = +0\,\kms) in LAB01
and redshifted by only +230\,\kms\ in LAB02.  These offsets are small
compared to those of Lyman break galaxies, which average +450\,\kms\
and extend to about +700\,\kms.  In LAB02, we detect \ion{C}{2}
$\lambda$1334 and \ion{Si}{2} $\lambda$1526 absorption lines, whose
blueward shifts of $\sim 200$\,\kms\ are consistent with the small
outflow implied by the redward shift of \lya.  We test and rule out the
simplest infall models and those outflow models with super/hyper-winds,
which require large outflow velocities.  Because of the unknown geometry
of the gas distribution and the possibility of multiple sources of \lya\
emission embedded in the blobs, a larger sample and more sophisticated
models are required to test more complex or a wider range of infall and
outflow scenarios.

\end{abstract}


%----------------------------------------------------------------------
\section{Introduction}

\lya\ nebulae, or ``blobs,'' are extended sources at $z$ $\sim$ 2--6
with typical \lya\ sizes of $\gtrsim$\,5\arcsec\ ($\gtrsim$\,50\,kpc)
and line luminosities of $L_{\rm{Ly\alpha}}\gtrsim10^{43}$ \unitcgslum\
\cite[e.g.,][]{Keel99, Steidel00, Francis01, Matsuda04, Matsuda10b,
Dey05, Saito06, Smith&Jarvis07, Hennawi09, Ouchi09, Prescott09, Yang09,
Yang10}.  Their number density and its large variance suggests that
blobs lie in massive ($M_{\rm halo}$ $\sim$ $10^{13}$\,\msun) dark
matter halos, which will evolve into those typical of rich clusters
today \citep{Yang10}.  Galaxies embedded within blobs are likely to
become brightest cluster galaxies.  Therefore, blobs are important sites
for studying the early interaction of galaxies with the surrounding
intergalactic medium (IGM).

This interaction is probably tied on some scale to the source of the
blobs' extended \lya\ emission, but the mechanism is poorly understood.
For example, the blob may represent emission from galactic-scale
outflows generated by star formation \citep{Taniguchi&Shioya00},
intense radiative feedback from AGNs \citep{Haiman&Rees01, Geach09},
or even cooling radiation from accreting gas \citep{Fardal01, Haiman00,
Dijkstra&Loeb09, Goerdt10}.

There are generally two ways to unravel the nature of the blobs.
The first approach is to study what powers the \lya\ emitting gas,
%% i.e., the energetics of \lya\ blobs, 
counting all ionizing sources within or around the extended \lya\ emission
and comparing that energy budget with the observed \lya\ luminosity.
Sometimes, one can identify possible energy sources, e.g., powerful AGNs
detected in X-rays \citep{Yang09, Geach09}.  Some potential sources may
even lie outside a blob \citep{Hennawi09}.  If no such source is found,
the remaining \lya\ luminosity may be attributable to cooling radiation
\cite[e.g.,][]{Nilsson06, Smith&Jarvis07, Smith08}.

An alternative path is to observe spectroscopically the kinematics of
the extended \lya-emitting gas.  Even addressing whether the gas is
falling into an embedded galaxy or outflowing into the IGM would be a
significant step forward, providing evidence for either gas accretion by
forming galaxies or for AGN/starburst-driven galactic winds, respectively.
Combining the two approaches by considering the blob's energetics and
kinematics simultaneously is ideal, a test of whether the bulk motions
of the gas are in fact coupled to the source of its illumination.
For example, gas infall is possible even in the presence of an AGN
\cite[e.g., ][]{Weidinger05, Humphrey07, Adams09}.  Cases in which blob
gas is both outflowing on small scales due to galactic processes and
infalling on larger scales from the extended IGM are not unlikely.

Unfortunately, the kinematic approach has been stymied by the limitations
of the \lya\ line, which, although bright compared with other lines, is
hard to interpret.  The ambiguity arises because \lya\ is a resonant line
and typically optically thick in the surrounding intergalactic medium.
As a result, studies even of the same blob's kinematics can disagree.
On one hand, \citet{Wilman05} argue that IFU spectra from a certain
blob are consistent with a simple model where the \lya\ emission is
absorbed by a foreground slab of neutral gas swept out by a galactic
scale outflow. On the other, \citet{Dijkstra06b} claim that the same data
can be explained by the infall of the surrounding intergalactic medium.
Worse, \citet{Verhamme06} comment that the same \lya\ profiles are most
consistent with static surrounding gas.

To distinguish among such possibilities requires a comparison of
the \lya\ line center with the center of an optically-thin line like
H$\alpha$\,$\lambda$6563 \cite[see also][]{Yang06}.  H$\alpha$ is a better
measure of the blob's systemic velocity, i.e., of the precise redshift,
because it is not seriously altered by radiative transfer effects and is
more concentrated about the galaxies in the blob's core.  Despite the
complexity of various blob models, infall models predict that the peak
of the optically thick \lya\ emission line must be blueshifted
%%
%% by a typical infall velocity (a few 100 km/s) 
%%
with respect to H$\alpha$, whereas outflows will redshift the \lya\
line \citep{Dijkstra06a}.  This diagnostic has been applied both to
local starburst galaxies \citep{Legrand97} and to Lyman break galaxies
(LBGs) at $z$ =1.5--2.5 \citep{Pettini01,Steidel04} and has provided
evidence for outflows driven by strong stellar winds.   An analysis in
a similar spirit, but using the \ion{H}{1} 21cm absorption line to mark
the systemic redshift of a high-$z$ radio galaxy, finds that the \lya\
line is blueshifted and thus that extended gas is inflowing to the AGN
\citep{Adams09}.  Previous spectroscopic observations of \lya\ blobs
\cite[e.g., at $z\gtrsim3$;][]{Wilman05, Matsuda06, Saito08} include
only \lya, precluding a direct comparison with a cleaner line's center
and muddling the interpretation of the kinematics.

To overcome this challenge, \citet{Yang09, Yang10} conduct blind surveys
for \lya\ blobs at $z\sim2.3$, a redshift at which important rest-frame
optical diagnostic lines (e.g., [\ion{O}{2}]\,$\lambda$3727,
[\ion{O}{3}]\,$\lambda$5007, H$\beta$\,$\lambda$4868,
H$\alpha$\,$\lambda$6563) fall in NIR windows and avoid bright OH sky
lines (or atmospheric absorption).  In this paper, we present optical and
NIR spectroscopy of the two brightest blobs, CDFS-LAB01 and CDFS-LAB02,
of the 25 in the \citet{Yang10} sample.  For each, we compare the \lya\
line profile with the center of the H$\alpha$ line and use existing
radiative transfer models \citep{Dijkstra06a, Verhamme06, Verhamme08}
to test whether the detected gas is infalling or outflowing.

In addition to the blob's systemic velocity, NIR spectroscopy can constrain
(1) the velocity dispersion and any ordered motion 
    like disk rotation of the brightest embedded galaxies,
(2) the presence of AGNs with rest-frame optical nebular line 
    diagnostics (e.g., \bptx), 
(3) the gas-phase metallicity, and 
(4) the star formation rate (SFR) directly from the H$\alpha$ flux, 
    allowing a comparison with that derived from the rest-frame UV.
Before now, none of these measurements has been available for \lya\ blobs.
In this paper, we focus on the kinematics of the gas in the blob and in
the embedded galaxies, deferring discussions of the AGN line diagnostics,
gas-phase metallicity, and SFRs to a future paper (Y.~Yang et al.\ 2011,
in preparation).

In \S\ref{sec:observation}, we review how we selected blobs from
our narrowband imaging surveys \citep{Yang10} and then describe the
optical/NIR spectroscopy of the first two blobs.  In \S\ref{sec:result},
we present the first results from the spectroscopic campaign.  We confirm
spectroscopically the blob redshift and discover multiple H$\alpha$
sources within each blob (\S\ref{sec:confirmation}).  We compare
the \lya\ profiles with the H$\alpha$ line centers to discriminate
between simple infall and outflow scenarios (\S\ref{sec:shift}).
In \S\ref{sec:absorption}, we describe the interstellar absorption lines
detected in LAB02 and compare the low ionization lines with H$\alpha$.
We estimate the velocity dispersions of three of the embedded galaxies
in \S\ref{sec:dynamical_mass}.
In \S\ref{sec:discussion}, we compare the \lya\ profiles and
\lya$-$H$\alpha$ offsets with existing radiative transfer models.  We also
discuss our results in the context of the clumpy circum-galactic gas
model proposed by \citet{Steidel10}.  \S\ref{sec:conclusion} summarizes
our conclusions.  Throughout this paper, we adopt cosmological parameters:
$H_0$ = 70\,${\rm km\,s^{-1}\ Mpc^{-1}}$, $\Omega_{\rm M}=0.3$,
and $\Omega_{\Lambda}=0.7$.

%----------------------------------------------------------------------
\section{Observations and Data Reduction}
\label{sec:observation}

%----------------------------------------------------------------------
%\input{./img/tile_hst_sinfo2.tex}
%\input{f01b.tex}
%----------------------------------------------------------------------

%----------------------------------------------------------------------
\begin{figure*}
%\epsscale{0.93}
\epsscale{1.15}
%\plotone{tile_hst_sinfo1.ps} 
%\plotone{tile_hst_sinfo2.ps} 
\plotone{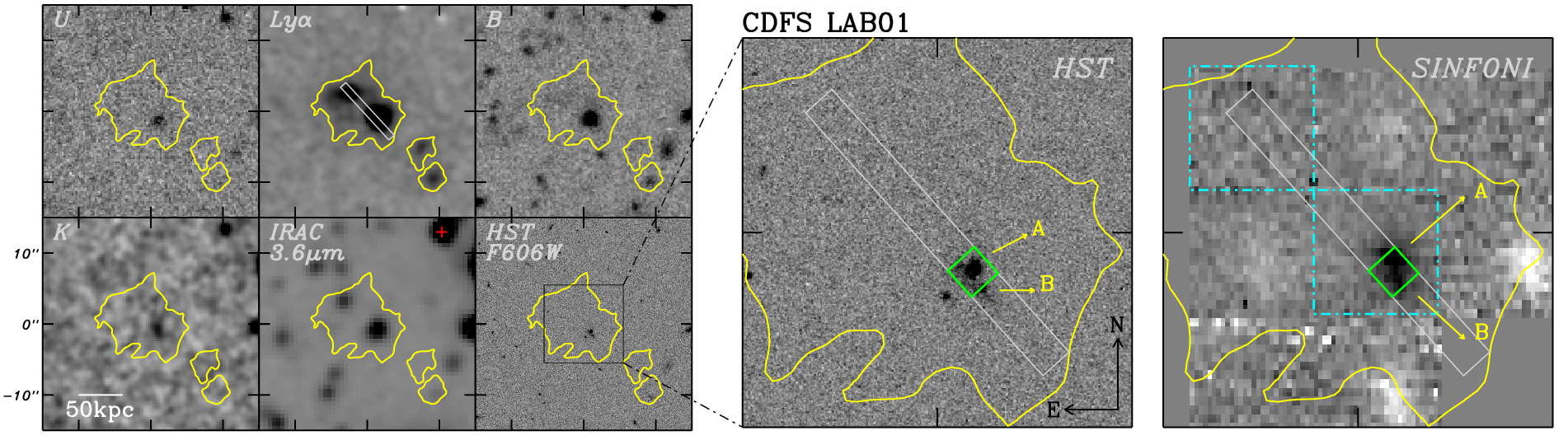} 
\plotone{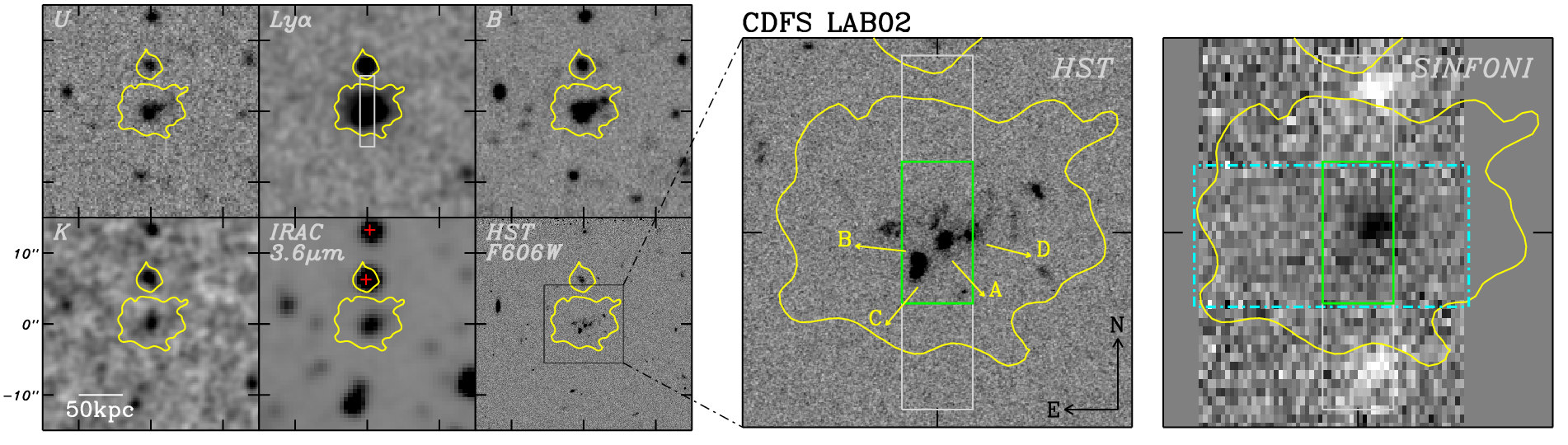} 
\caption{
({Left}) Images of \lya\ blobs at various wavelengths: {\sl U},
continuum-subtracted \lya\ line, {\sl B}, {\sl K}, {\sl Spitzer} IRAC
3.6\micron, and {\sl HST} F606W images.  Contours represent a surface
brightness of 4$\times$10$^{-18}$ \unitcgssb\ from our CTIO-4m narrowband
imaging. Ticks represent 10\arcsec\ (82 physical kpc) intervals. The
red crosses in the IRAC 3.6\micron\ image indicate X-ray sources from
the CDFS and E-CDFS catalogs.
({Right}) High-resolution {\sl HST} and SINFONI H$\alpha$ images collapsed
in the wavelength direction at $\lambda$ $\sim$ 2.175\micron.
Here, we show the SINFONI observation taken in the best seeing for
LAB01 (1.3 hrs with 0\farcs3 in K-band).  For LAB02, the seeing was
$\sim$1\farcs2.
SINFONI reveals two and three relatively bright H$\alpha$ sources
within LAB01 and LAB02, respectively. These sources roughly correspond
to embedded galaxies or galaxy fragments seen in the {\sl HST} images
(labeled as A, B, C and D).
The gray and green boxes represent the MagE slit positions
and the extraction windows for the spectra shown in Figure
\ref{fig:line_comparison}, respectively.  In the SINFONI image, we show
the final FOV resulting from the dither patterns (dot-dashed lines).
The negative images outside the FOV are the ghost images due to sky
subtraction.
}
\label{fig:image}
\end{figure*}
%----------------------------------------------------------------------

%----------------------------------------------------------------------

\subsection{Sample}

We observe the two brightest \lya\ blobs from the \citet{Yang10}
sample, which consists of 25 blobs discovered in four different survey
fields. These two blobs lie in the Extended Chandra Deep Field South
(E-CDFS) and were found via deep imaging with the CTIO-4m MOSAIC--II
camera and a custom narrowband filter (\nb403).  This filter has
a central wavelength of $\lambda_c \approx 4030$\AA, designed for
selecting \lya-emitting sources at $z\approx2.3$, and a band-width of
$\Delta\lambda_{\rm FWHM} \approx 45$\AA, providing a line of sight
(hereafter LOS) depth of $\Delta z \simeq 0.037$ that corresponds to
46.8 comoving Mpc.
The blob selection criteria are: EW$_{\rm obs}$ $>$ 100\AA\ and isophotal
area $A_{\rm iso}$ $>$ 10\,\sq\arcsec\ above the surface brightness
threshold of 5.5 $\times$ 10$^{-18}$ \unitcgssb.  We show the images of
the two blobs (CDFS-LAB01 and CDFS-LAB02) in Figure \ref{fig:image}. We
refer readers to \citet{Yang10} for details of the sample selection.

%--------------------------------------------------
\subsection{Optical Spectroscopy}

We obtained high resolution optical spectra of the two \lya\ blobs using
the Magellan Echellette Spectrograph \cite[MagE;][]{Marshall08} on the
Magellan Clay 6.5m telescope.  MagE provides a spectral resolution of $R$
$\sim$ 4100 for a 1\arcsec\ slit over a wide wavelength range of 3000\AA\
-- 10000\AA.
While the slit length is small (10\arcsec) compared to typical longslits,
the excellent blue sensitivity around 4000\AA\ is ideal  for obtaining
the \lya\ line profile.  Observations were carried out on UT 2008 July
26 and October 1, and 2009 February 17.  For LAB01, we first acquired a
low resolution spectrum in 2008 July 26 to spectroscopically confirm
the redshift prior to the NIR spectroscopic run.   Later, we obtained
the $R\sim4100$ spectrum with a slit orientation of P.A.=42\degr\ as
shown in Figure \ref{fig:image} (\lya\ panel).
%%
%% two slit orientations and widths (P.A.=$-42$\degr\ with 1\arcsec\
%% and P.A.=0\degr\ with 2\arcsec) as shown in Figure \ref{fig:image}
%% (\lya\ panel).
%%
For LAB02, we used a 2\arcsec\ slit and a P.A.=0\degr\ to obtain a $R$
$\sim$ 2000 spectrum.
The sky condition was clear, and the seeing ranged from 0.9 to 1.3\arcsec.
The individual exposure times varied from 20 to 40min, depending on the
variability of the seeing.  Total exposure times were 2.5 and 2.0 hrs
for LAB01 and LAB02, respectively.
For accurate wavelength calibration, we took ThAr lamp frames right before
and after the science exposures and at the same telescope pointing to
eliminate systematic errors introduced by instrument flexure.

We reduce the MagE data using the Daniel Kelson's Carnegie-Python
Package \citep{Kelson00b,Kelson03}.  The frames are overscan-corrected,
bias-subtracted, and flat-fielded with Xenon-flash and quartz
(incandescent) lamps.  Using night sky lines and ThAr lamp spectra, we
estimate the uncertainties in the wavelength solution to be $\sim$0.03\AA\
(0.1 pixel).  The sky background is then subtracted in 2D using the
method described in \citet{Kelson03}, and the frames are corrected for
spatial distortions. We combine the rectified 2D spectra and extract the
1D spectra using IDL routines. We also re-reduce the data with the MagE
Spectral Extractor package \cite[{\tt MASE};][]{Bochanski09} and confirm
that the results are consistent.  The spectra are flux-calibrated with
2--3 standard stars each night.
Due to the limited spatial coverage of the slit and the relatively short
exposure times, our spectra are dominated by the brightest spots within
the blobs, which coincide with the galaxies detected in the {\sl HST}
rest-frame UV images.  Future IFU observations should be helpful in
constraining how the \lya\ profiles vary over the blobs.

\subsection{Near Infrared Spectroscopy}

%----------------------------------------------------------------------
%\input{./img/plot_spec1d.tex}
%\input{f02.tex}
%----------------------------------------------------------------------

%----------------------------------------------------------------------
\begin{figure*}
%\epsscale{1.0}
\epsscale{1.1}
%\plotone{plot_spec1d.ps}
\plotone{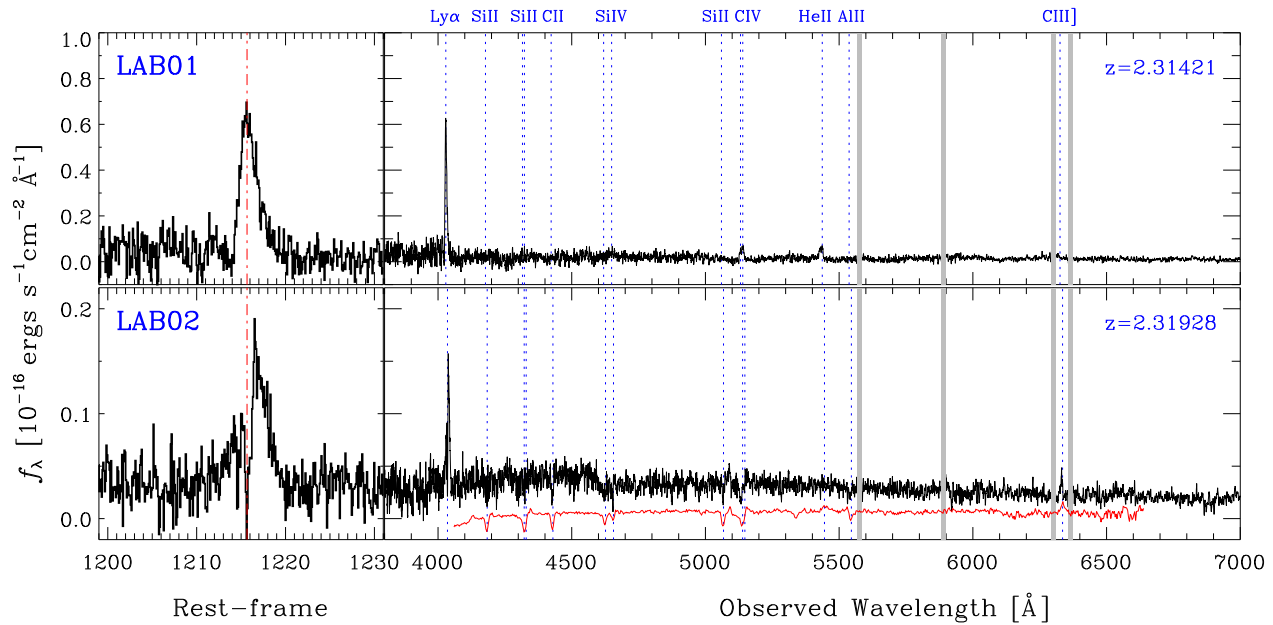}
\caption{
Integrated 1D \lya\ spectra for LAB01 and LAB02 at $z\simeq2.31$.
The spectra were extracted from apertures of 1\arcsec$\times$2\farcs4
(LAB01) and 2\arcsec$\times$2\farcs4 (LAB02).
({Right}) MagE spectra in the observed wavelength frame from 3500\AA\ --
7000\AA. Vertical dot-dashed lines indicate the location of important UV
emission/absorption lines. The gray bars indicate the masked region due to
strong sky lines. The spectra are boxcar-smoothed with 5 spectral pixels.
In addition to strong \lya\ emission at $\sim$4030\AA, LAB01 shows
\ion{C}{4} $\lambda$1549 and \ion{He}{2} $\lambda$1640 lines.  In LAB02,
UV continuum from the embedded galaxies, possibly galaxies A and B+C,
is detected. For comparison, we show the composite LBG spectrum from
\citet{Shapley03}.
({Left}) Close-up of \lya\ profiles in the rest-frame.  We convert the
observed spectra into rest-frame wavelength using the redshifts determined
from their H$\alpha$ lines ($z$ = 2.31421 and 2.31928).
}
\label{fig:spec1d}
\end{figure*}
%----------------------------------------------------------------------

%----------------------------------------------------------------------

We carried out NIR Integral Field Unit (IFU) spectroscopy of the two \lya\
blobs using SINFONI \citep{Eisenhauer03, Bonnet04} on the VLT UT4
telescope in visitor mode between UT 2009 January 30 and February 4.
To measure the redshifted H$\alpha$ ($\lambda_{\rm obs}$ $\sim$
2.175\micron) line, we used the  {\sl K}-band grating, which covers
1.95\micron\ -- 2.45\micron\ with a spectral resolution of $R$
$\simeq$ 4000 ($\Delta v$ $\simeq$ 75\,\kms) and dispersion of
2.5\,\AA\,pixel$^{-1}$.  Because the blobs are spatially extended
over 5\arcsec--10\arcsec\ and include multiple sources, we employed
the seeing-limited mode (i.e., no adaptive optics), which provided
the largest FOV (8\arcsec$\times$8\arcsec) and a plate scale of
0\farcs25\,pixel$^{-1}$ for the spatial resolution elements (spaxels).
%%
%% The seeings measured in the optical from the DIMM seeing monitor
%% ranges from 0.6\arcsec\ to 1.2\arcsec.
%%
In each observing block, we observed reference stars, which we used to
blind offset to the blobs as well as to monitor the telescope pointing
accuracy and the seeing variations.  The {\sl K}-band seeing ranged from
0\farcs3 to 1\farcs2, and the sky condition was clear.
To maximize the on-source integration time, we adopted an ``on-source
dithering'' strategy in which the science targets were always kept
within the FOV but at different detector positions.  For LAB01, we
used a three-point dithering scheme that moves among the pointings
by $\sim$4\arcsec\ to south or west, providing a usable FOV of two
4\arcsec$\times$4\arcsec\ squares (see the SINFONI panel in Figure
\ref{fig:image}).
For LAB02, we employed a two-point dither pattern in the north-south
direction, resulting in a 8\arcsec$\times$4\arcsec\ effective FOV
(dot-dashed line in Figure \ref{fig:image}).  The total integration times
for LAB01 and LAB02 are 6 hrs and 0.5 hrs, respectively, which consist
of 5 or 10 min individual exposures depending on the sky condition.
For the telluric correction and flux calibration, we observed four B or
G stars every night before or after the science targets with wide range
of airmass.

%----------------------------------------------------------------------
%\input{./img/extract_line_bpt_lab02.tex}
%\input{f03b.tex}
%----------------------------------------------------------------------

%----------------------------------------------------------------------
\begin{figure*}
\epsscale{1.1}
%\plottwo{extract_line_bpt_lab01.ps}{extract_line_bpt_lab02.ps}
\plottwo{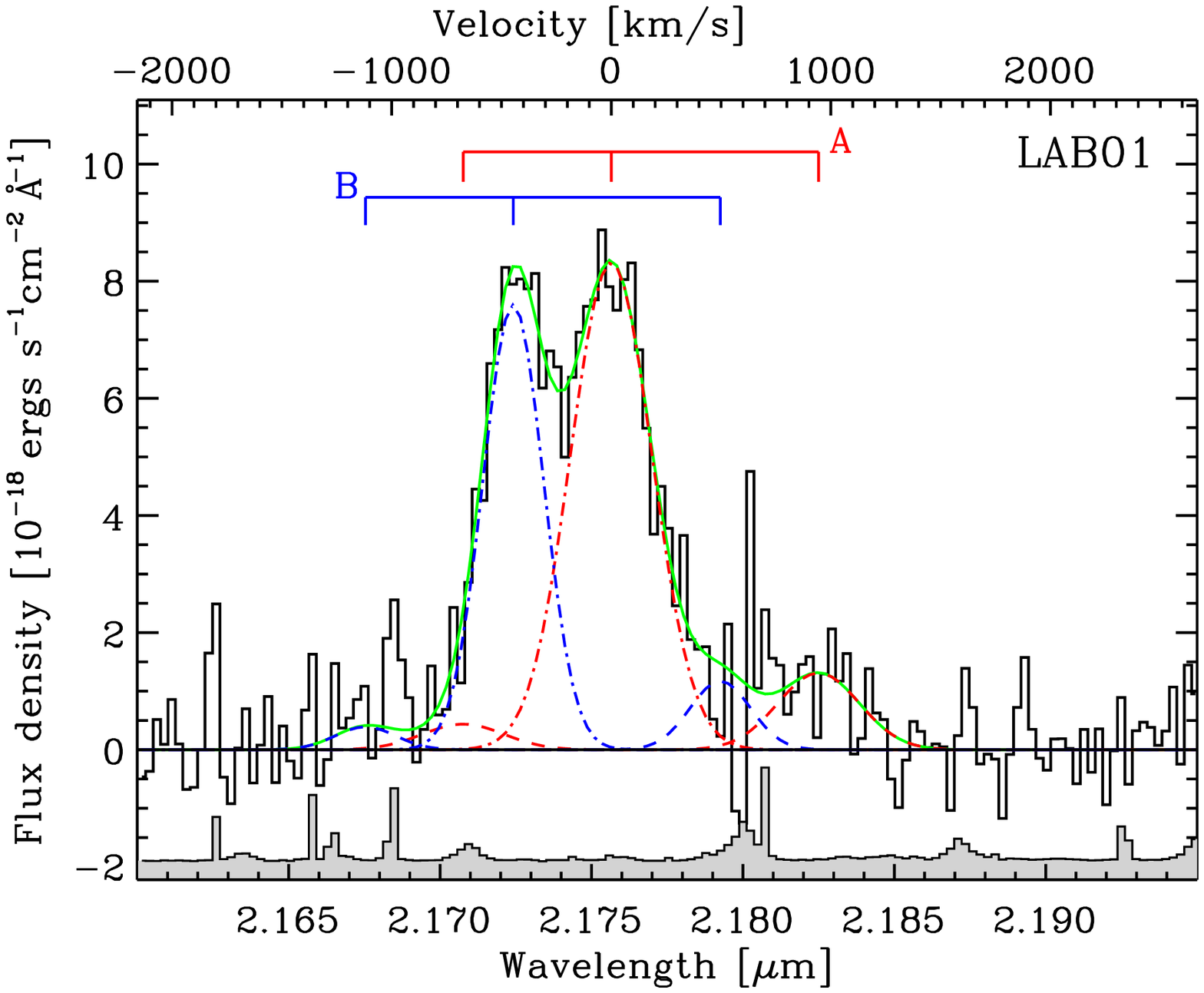}{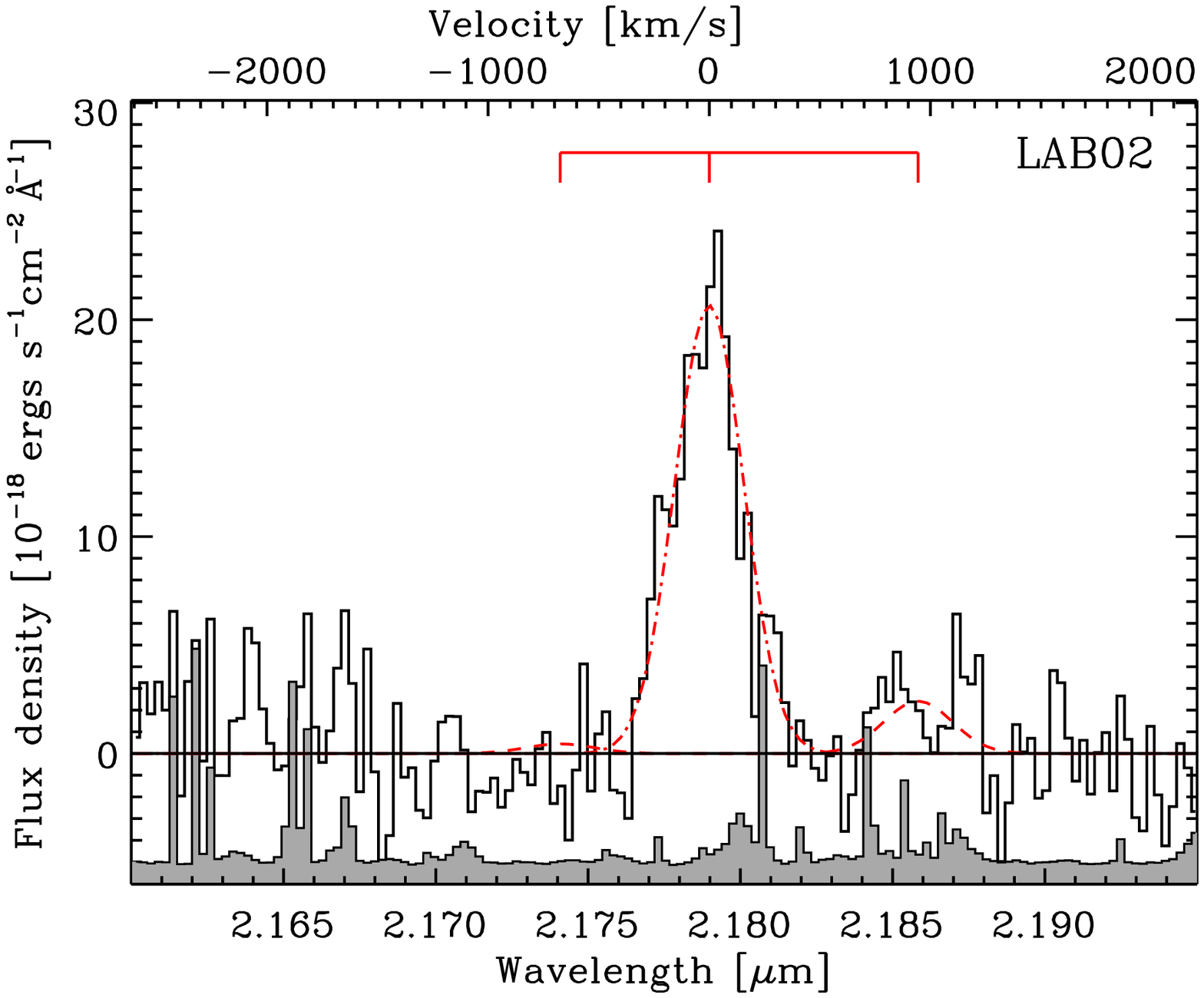}
\caption{
Integrated 1D spectra of the H$\alpha$ and [\ion{N}{2}]
$\lambda\lambda$6549,6583 lines extracted from a 2\arcsec$\times$2\arcsec\
aperture for LAB01 and LAB02.
Note that to maximize the S/N, we adopt different aperture sizes here
than those used for \lya\ in Figure \ref{fig:spec1d}.
The dashed and dot-dashed lines represent Gaussian fits to [\ion{N}{2}]
and H$\alpha$, respectively.  The gray histograms in the lower panels
show the uncertainties.
For LAB01, two sets of H$\alpha$+[\ion{N}{2}] lines from galaxies A and B
are indicated.  We use all available data (6\,hrs) taken in various seeing
(0\farcs3 -- 1\farcs2). These two galaxies have a velocity difference
$\Delta v$\,(A--B) = 440\,\kms.  In our subsequent analyses, we compare
only the \lya\ and H$\alpha$ lines extracted along the line of sight to
galaxy A.
For LAB02, we are not able to reliably extract individual profiles
from galaxies B and C, because of the low S/N. Thus we fit only one
H$\alpha$+[\ion{N}{2}] set to the LAB02 spectrum, which is dominated by
galaxy A.
}
\label{fig:linefit}
\end{figure*}
%----------------------------------------------------------------------

%----------------------------------------------------------------------

We reduce the SINFONI data using the ESO pipeline. Dark current and
sky background are removed from each science frame by subtracting the
pseudo-sky frame, which is constructed from the average of the two science
frames bracketing each science frame.  Then, we flat-field the data and
correct for bad pixels.
Because we will compare the velocity centers of the \lya\ and H$\alpha$
lines, accurate wavelength calibration is critical.  We compare the
wavelength solution obtained from the OH sky lines in the science frames
to that from the daytime arc lamps.  The arc lamp wavelength calibration
error arising from instrument flexure can be as large as 1.5\AA, depending
on the airmass (which ranges up to $\sim$2.0).  Therefore, we adopt the OH
sky line wavelength calibration, whose error is typically $\sim$ 0.1\AA\
(0.04 pixel) for each science frame.
Using this wavelength solution, we constructed 3D data cubes from the
2D spectra after correcting spatial distortion.
Because it is not possible to identify/centroid the sources in individual
exposures, we align them according to the dither offsets within each
observing block (OB).  The spatial offsets between OBs are calculated
from the PSF calibration frames and are typically small ($\sim$ 0\farcs1).
When combining the data cubes, we adopt an iterative $\sigma$-clipping
algorithm to reject possible outliers and produce the ``sigma cubes''
that represent the standard deviation of the adopted pixels.  Because
no stellar continuum is detected, 
we subtract the median sky values from each wavelength plane to avoid the
strong variation of global sky background.  Finally, we flux-calibrate
the cubes using the broad-band magnitudes of standard stars.
Atmospheric absorption is minimal at H$\alpha$ due to our choice of
survey redshift, so we do not apply a telluric correction.

% However, note that the effect of imperfect telluric correction is
% limited due to our selection of survey redshifts where atmospheric
% absorption and OH night sky emission is minimal.

%----------------------------------------------------------------------
\section{Results}

\label{sec:result}

\subsection{Redshift Confirmation with \lya\ and H$\alpha$ Lines}
\label{sec:confirmation}

We spectroscopically confirm the redshifts ($z$ $\simeq$ 2.31 -- 2.32)
of the two \lya\ blobs using both the \lya\ and H$\alpha$ emission lines.
Figure \ref{fig:image} shows the two \lya\ blobs at various wavelengths:
{\sl U}, continuum-subtracted \lya\ line, {\sl B}, {\sl K}, {\sl Spitzer}
IRAC 3.6\micron, and {\sl HST} F606W images overlain with the \lya\
contours ({left six panels}).  The archival ground-based images ({\sl
U, B}, and {\sl K}), the {\sl Spitzer} IRAC images, and the {\sl HST}
images are obtained from the Multiwavelength Survey by Yale-Chile
\cite[MUSYC;][]{Gawiser06a}, the Spitzer IRAC/MUSYC Public Legacy in
E-CDFS survey \cite[SIMPLE;][]{Damen10}, and the Galaxy Evolution from
Morphology and SEDs \cite[GEMS;][]{Rix04} survey, respectively.  In the
right panels, we show the enlarged {\sl HST} images and SINFONI H$\alpha$
intensity maps that are collapsed in the wavelength direction centered
at $\lambda$ $\simeq$ 2.1746\micron\ ($\Delta\lambda$ $\simeq$ 66\AA)
and 2.1788\micron\ ($\Delta\lambda$ $\simeq$ 42\AA) for the two blobs,
respectively.  For the SINFONI image of LAB01, we show only data taken
in the best seeing ($\sim$ 0\farcs3) to be comparable with the {\sl HST}
resolution.

%----------------------------------------------------------
%         from        to          center      dlambda
% LAB01   2.1713350   2.1779500   2.1746425   -66.150001
% LAB02   2.1769700   2.1811350   2.1788320   -41.650000
%----------------------------------------------------------

Interestingly, in the {\sl HST} rest-frame {\sl UV} images, both \lya\
blobs include multiple galaxies and/or fragments, some of which appear
to be interacting \cite[see also][]{Colbert06}.  A few are also detected
in H$\alpha$, indicating that they lie at the same redshift.
In \citet{Yang10}, we found that the number density and the field-to-field
variance of \lya\ blobs are consistent with their occupying massive dark
matter halos ($\sim$ $10^{13}$\,\msun) that probably evolve into those
of typical galaxy clusters today.  Thus, while some clumps/fragments
could be part of a one large galaxy \cite[e.g.,][]{Labbe03,Elmegreen09},
we speculate that others may merge and evolve into present-day brightest
cluster galaxies.

%----------------------------------------------------------------------
%\input{./img/collapse_cube_LAB02.tex}
%\input{f04b.tex}
%----------------------------------------------------------------------

%----------------------------------------------------------------------
\begin{figure*}
\epsscale{1.0}
\epsscale{1.0}
\centering
\includegraphics[height=0.39\textwidth]{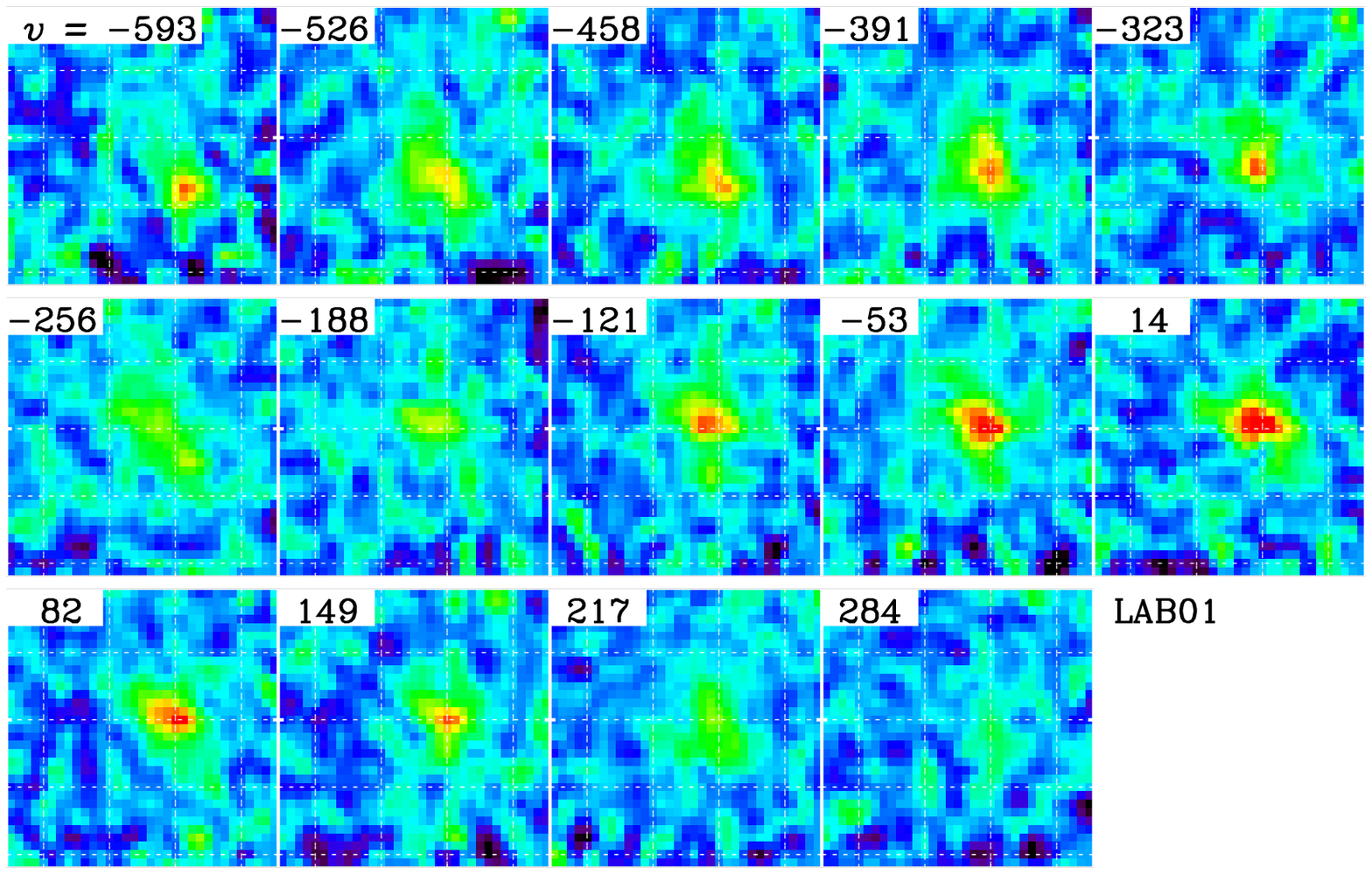}
\includegraphics[height=0.39\textwidth]{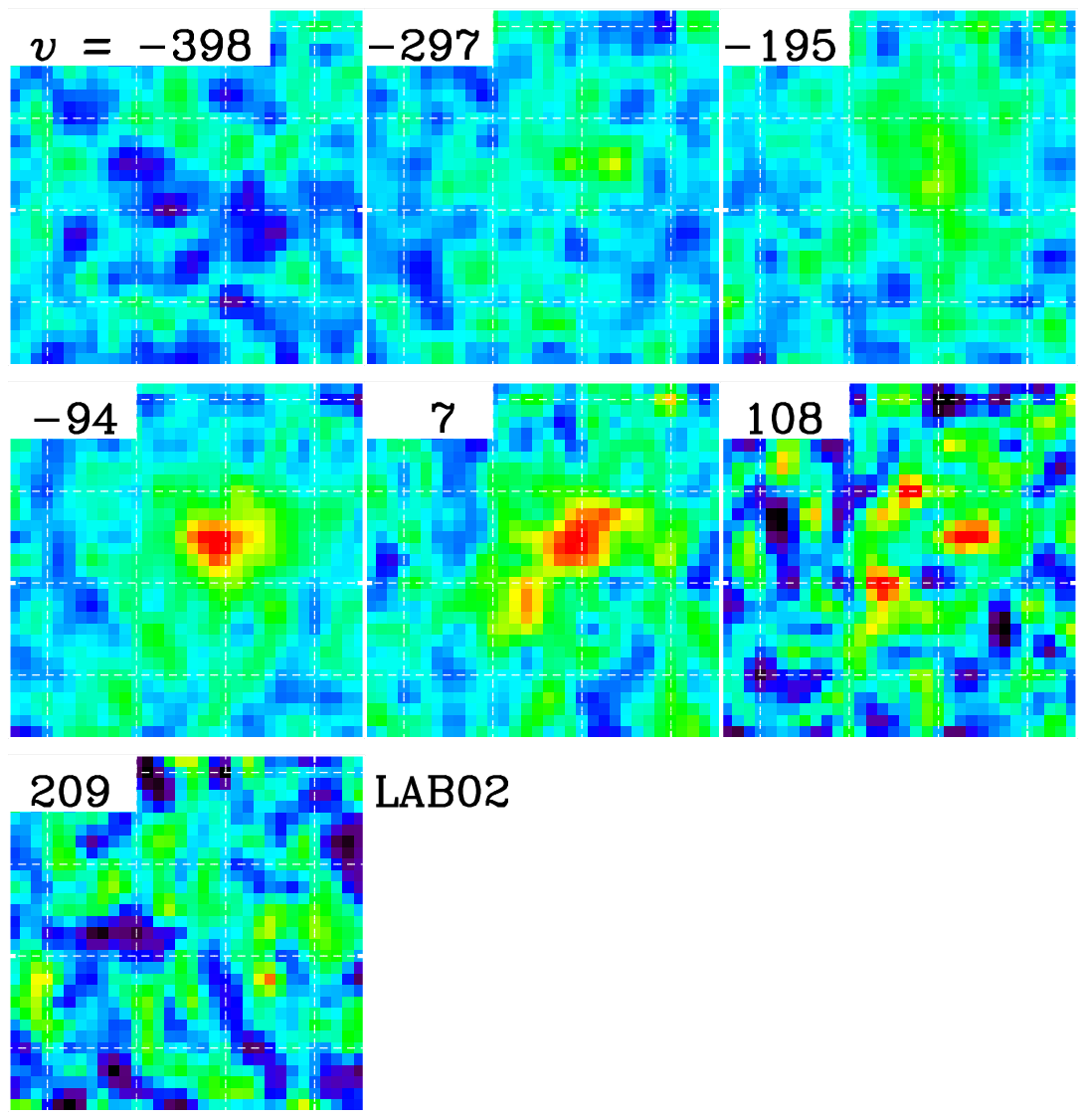}
\caption{
H$\alpha$ channel map for LAB01 and LAB02, respectively.  The grids
represent 1\arcsec\ intervals, and each channel is centered at the
velocity noted at the upper left corner of each panel.  Each intensity
map is smoothed with a Gaussian kernel of FWHM=0\farcs25.  In LAB01,
there are two components that are spatially and kinematically distinct,
which we associate with galaxy A and B in the {\sl HST} images.  In LAB02,
the H$\alpha$ emission is dominated by galaxy A.  There are two
weak features that might correspond to galaxies B+C and D;
because of the low S/N, we are not able to extract spectra for them.
The last two channels of LAB02 are strongly affected by an OH sky line.
}
\label{fig:channel}
\end{figure*}
%----------------------------------------------------------------------

%----------------------------------------------------------------------

Figure \ref{fig:spec1d} shows the 1D \lya\ spectra extracted from
1\arcsec$\times$2\farcs4 and 2\arcsec$\times$2\farcs4 MagE slits for
LAB01 and LAB02, respectively.
In addition to strong \lya\ emission, LAB01 also has \ion{C}{4}
$\lambda$1549 and \ion{He}{2} $\lambda$1640 emission lines, implying the
presence of hard ionizing sources \cite[e.g,][]{Prescott09, Scarlata09}.
We defer analysis of these lines and their relative strengths to a
future paper (Y.~Yang et al. 2011, in preparation), which focuses
on the multi-wavelength properties of the full blob survey and the
implications for the sources of blob emission.  LAB02 has several
interstellar absorption lines that we use in \S\ref{sec:absorption}
to help constrain the kinematics of gas around the embedded galaxies.

In contrast to the optically thick and spatially extended \lya\ emission,
we can use the optically thin, more spatially concentrated H$\alpha$
emission to make an independent determination of the blob redshift
from the galaxies embedded in the centers of the blobs.  We extract
the H$\alpha$ spectra from a 2\arcsec$\times$2\arcsec\ aperture in the
SINFONI cubes. 
Figure \ref{fig:linefit} shows the resulting spatially integrated
H$\alpha$ profiles within the blobs. The LAB01 H$\alpha$ spectrum includes
all 6\,hr exposures combined to obtain higher S/N and total flux.
Note that whenever we show individual \lya\ or H$\alpha$ spectra (Figures
\ref{fig:spec1d} and \ref{fig:linefit}), we adopt different aperture sizes
that maximize the S/N.  On the other hand, we use the same aperture when
comparing the \lya\ and H$\alpha$ lines directly (\S\ref{sec:shift}).

\setcounter{footnote}{0}

For LAB01, we focus on the brighter SW component of the blob.  The
excellent seeing of our SINFONI observation ($\sim$ 0.3\arcsec) allows us
to spatially resolve two embedded galaxies (A and B) in the SW part of the
blob (hereafter LAB01A and LAB01B; Fig.~\ref{fig:image})\footnote{While
it is possible that these two components could be \ion{H}{2} regions
or star-forming clumps in a single galaxy, it is unlikely given that
their projected separation is $\sim$5\,kpc and the velocity offset is
$\sim$440\,\kms.}.
These two galaxies are also
kinematically separated as shown in the SINFONI H$\alpha$ channel
map (Figure \ref{fig:channel}), a series of H$\alpha$ intensity maps
constructed by stepping through the 3D SINFONI data cubes in increments
of 70\,\kms. Galaxy B shows an apparent velocity shear of $\sim$ 250
\kms, suggesting disk rotation, while galaxy A has no obvious velocity
structure.
In their spatially integrated H$\alpha$ spectra, these two galaxies have
the nearly same redshift, $z^A_{\rm H\alpha}$ = 2.31421 $\pm$ 0.0001 and
$z^B_{\rm H\alpha}$ = 2.30928 $\pm$ 0.0001, a separation of $\Delta z$
= 0.0049 or $\Delta v$(A--B) = 440\,\kms.
This relative velocity is a poor estimate of the velocity
dispersion $\sigma$ of the halo in which these galaxies are embedded.
Nevertheless, it is consistent with the $\sigma$ expected for a massive,
$\sim$$10^{13}$\,\msun\ halo.

LAB02 also contains several continuum sources in the {\sl HST} UV images
(labeled as A--D).  At least one of the brightest sources (LAB02A) is
detected in the SINFONI H$\alpha$ image.  With SINFONI, we marginally
detect the two sources (B+C) blended in the {\sl HST} image.  It is also
possible that galaxy D is blended with the much brighter galaxy A in
the IFU data.  We are not able to reliably extract H$\alpha$ profiles
from galaxies (B+C) and D because of the shallow exposure (30 min) and
insufficient spatial resolution.  The integrated spectrum of LAB02 is
dominated by the brightest galaxy A, for which we measure the $z_{\rm
H\alpha}$ = 2.31928 $\pm$ 0.0001. We summarize the properties of 
H$\alpha$ lines in Table \ref{tab:line_properties}.

While it is possible that the H$\alpha$ detections include some extended
gas, the centers of the emission coincide with the galaxies resolved
in the {\sl HST} image and are likely to originate there.  Therefore,
in the following section, we compare the \lya\ and H$\alpha$ profiles
along the LOS toward these embedded galaxies.
There are small astrometry offsets ($\sim$0.2\arcsec) between {\sl
HST} UV images and SINFONI H$\alpha$ maps (right panels in Figure
\ref{fig:image}). While it is likely that this mismatch arises from the
astrometric calibrations of SINFONI, which rely entirely on blind
offsets, we extract the \lya\ and H$\alpha$ spectra from the same
aperture, accepting the current astrometric solution.  Changing the
relative position of extraction aperture (green boxes) for \lya\
and H$\alpha$ does not affect any of the conclusions in this paper.

We show the selected extraction apertures (small green boxes) in
Figure \ref{fig:image}.  In LAB01, we extract the spectra along the
LOS toward galaxy A, because the peak of \lya\ map coincides with
galaxy A.  While galaxies A and B are well-resolved in the H$\alpha$
image, the spectrum extracted from the sight-line to LAB01A (green box
in Fig.~\ref{fig:image}) is slightly contaminated with light from galaxy
B.  Therefore, we simultaneously fit the profile with a total of six
components (three for each galaxy) and remove the galaxy B components when
comparing the H$\alpha$ and \lya\ line centers in the following section.

%----------------------------------------------------------------------
\subsection{Comparison Between \lya\ and H$\alpha$ Lines}
\label{sec:shift}

The \lya\ profile emerging from the blob should be complicated, a
mixture of the bulk motions of the gas and geometry.  For the details
of \lya\ radiative transfer, we refer readers to the literature
\cite[e.g.,][]{Ahn01, Ahn02, Dijkstra06a, Verhamme06, Kollmeier10,
Faucher-Giguere10}.  In this section, we compare the \lya\ and
H$\alpha$ lines to constrain the gas kinematics within the blobs.
More specifically, we measure the velocity shift of the \lya\ line
relative to the H$\alpha$ line center to distinguish between any infall
and outflow of the IGM relative to the blob's systemic velocity.  Once we
measure this first-order kinematic diagnostic (i.e., any blueshift or
redshift of the \lya\ line),  we discuss the detailed \lya\ and H$\alpha$
profiles in \S\ref{sec:discussion}, comparing them with radiative transfer
calculations to determine the velocity of the dominant bulk motion.

%----------------------------------------------------------------------
%\input{./img/profile_lab02A.tex}
%\input{f05b.tex}
%----------------------------------------------------------------------

%----------------------------------------------------------------------
\begin{figure*}
\epsscale{1.15}
%\plottwo{profile_lab01A.ps}{profile_lab02A.ps}
\plottwo{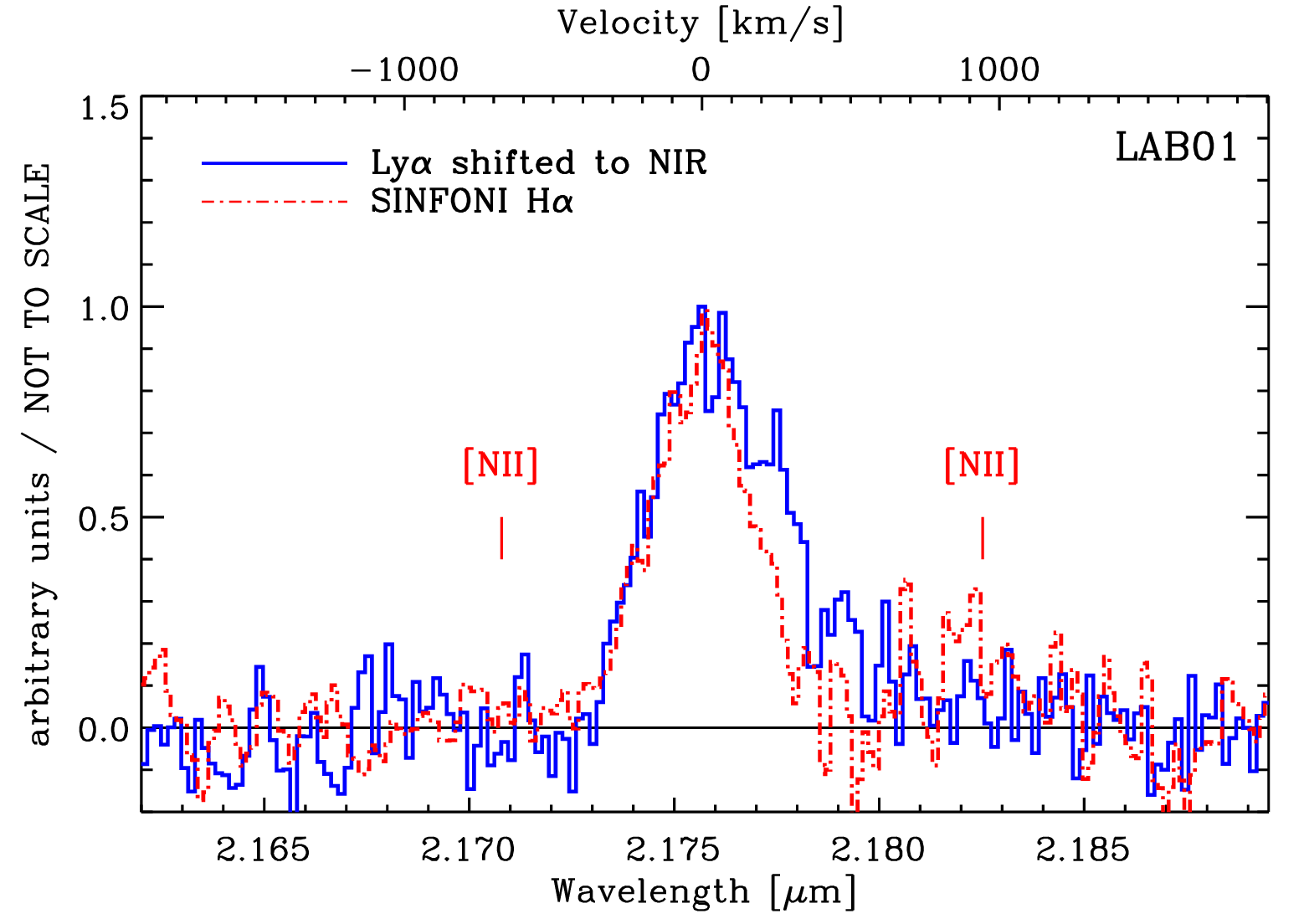}{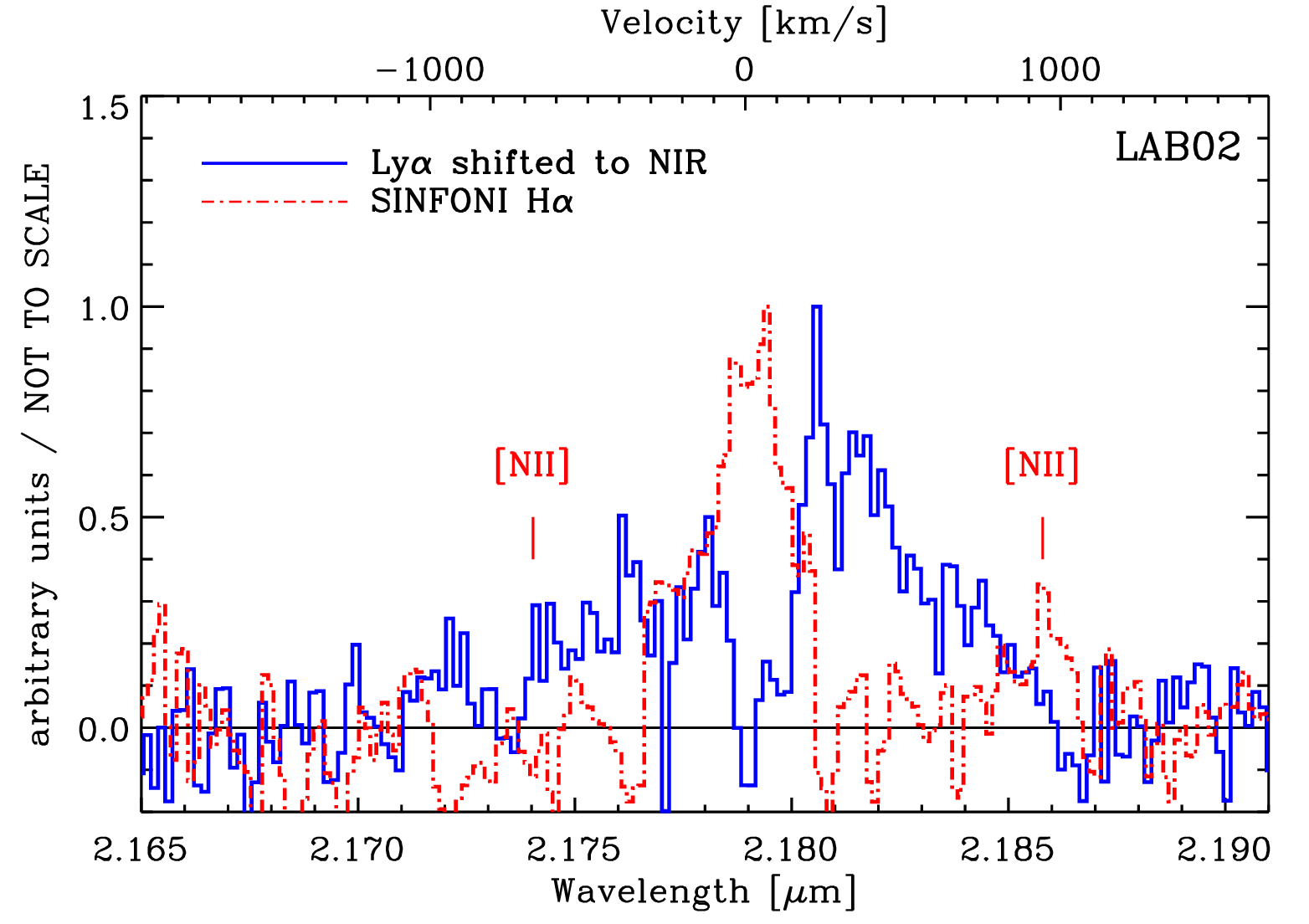}
%\plotone{f05a.ps}
%\plotone{f05b.ps}
\caption{
Comparison between optically thick \lya\ line (blue, solid) and optically
thin H$\alpha$ line (red, dashed) for the two \lya\ blobs.  For both
LAB01 and LAB02, we are comparing the \lya\ profiles with H$\alpha$
lines extracted from along the LOS toward galaxy A.
Note that both \lya\ and H$\alpha$ are extracted from the same aperture
({\it green boxes} in Figure \ref{fig:image}) for appropriate comparison.
The \lya\ profiles are transformed to the observed NIR wavelength
(SINFONI). For reference, we show the velocity frame obtained from
the H$\alpha$ redshift on the top $x$-axis.
%%
%% The vertical arrows in LAB01 indicate possible multiple peaks as 
%% discussed in \S\ref{sec:discussion}.
%%
In both blobs, the \lya\ lines are broader than the H$\alpha$ lines,
implying that \lya\ is resonantly scattered by an optically thick
medium.  The \lya\ peaks either agree with or are redshifted against
their H$\alpha$ line center (\dvlya\ = 0\,\kms\ in LAB01 and 230\,\kms\
in LAB02), suggesting no inflow and a possible outflow, respectively.
}
\label{fig:line_comparison}
\end{figure*}
%----------------------------------------------------------------------

%----------------------------------------------------------------------

To illustrate what we expect from this experiment, we first consider the
simplest geometric case:  a galaxy surrounded by a spherically symmetric
gas cloud that is static, collapsing, or expanding \citep{Dijkstra06a,
Verhamme06}.
Here we assume that the H$\alpha$ line arises from the embedded galaxy,
as we do not detect extended H$\alpha$ emission, and that the \lya\ line
originates either in the same region of the galaxy before eventually escaping
the surrounding cloud or in the cloud itself.
Thus this blob will be observed as a point source in H$\alpha$, but as an
extended source in \lya.  Note that the \lya\ and H$\alpha$ spectroscopy
is along the LOS toward the central galaxy.  While H$\alpha$ from the
galaxy can escape the surrounding gas without suffering absorption, \lya\
photons from the galaxy and from gas along the LOS will be resonantly
scattered many times until their frequencies are right for escaping the
cloud, a condition dependent on the bulk motions of gas along the LOS.

Because the cloud is optically thick to the \lya\ line, \lya\ photons
on the Doppler wings should escape the cloud through random scatterings
in the frequency domain.  If the surrounding gas is static, the \lya\
line will have a double-peaked profile, while the optically thin
H$\alpha$ photons will escape without any radiative transfer effects.
If the cloud is collapsing, \lya\ photons within the cloud on the red
side of double-peaked profile will see higher optical depth due to the
LOS infalling gas, and the red peak will be depressed.  If the
gas is outflowing, the blue side of the profile will be more diminished.

Therefore, if there is gas infall, the \lya\ profile will be asymmetric
and blueshifted against the H$\alpha$ line due to radiative transfer
in the optically thick medium, but the corresponding H$\alpha$ line
profile should be symmetric.  For outflowing gas, the \lya\ will be
redshifted against the symmetric H$\alpha$ line.  The amount of shift
depends on various parameters such as infall/outflow velocity and optical
depth.
Note that one cannot determine whether the \lya\ line is redshifted
or blueshifted against the background velocity field unless there is a
optically thin reference line, i.e., H$\alpha$.

The actual geometry of the system is likely to be more complicated
(\S\ref{sec:discussion}).  Yet the blue- or redshift of the \lya\
photons relative to the systemic velocity will occur as they make their
way out of the system, as long as they encounter optically-thick gas that
is infalling or outflowing along the LOS \citep{Dijkstra06a}.  In other
words, we are studying the bulk motion of gas along the pencil beam toward
the embedded galaxies when we compare the \lya\ and H$\alpha$ profiles
extracted from this sight-line.  In principle, it is even possible to
compare the velocity offsets between extended \lya\ and {\it extended}
H$\alpha$ lines in two dimensions (i.e., H$\alpha$ and \lya\ screens)
with much deeper H$\alpha$ spectroscopy and optical IFU data.

In summary, the basic assumptions for our simple test using the velocity
offset between the \lya\ and H$\alpha$ lines are:
%---------------------------------
\begin{enumerate} %\begin{itemize}
\item \lya\ should go through an optically thick column of gas (either
infalling or outflowing) to escape.
%% while the detailed distribution of this gas is unknown (e.g, clumpy
%% or continuous).

\item H$\alpha$ represents the systemic velocity of the entire system,
and gas is moving (infalling or outflowing) relative to this center.
%% while a few or several galaxies are observed in blobs.
\end{enumerate} %\end{itemize}
%---------------------------------

With these assumptions in mind, we now consider the \lya\ and
H$\alpha$ line profiles for our two \lya\ blobs.  We then examine these
assumptions further and discuss how to overcome their limitations in
\S\ref{sec:discussion}.

We show the \lya\ and H$\alpha$  profiles for the two \lya\ blobs in
Figure \ref{fig:line_comparison}.  We shift each blob's \lya\ profile
into its H$\alpha$ frame using
$\lambda^{\rm H\alpha}_{\rm NIR}$ = 
$\lambda^{\rm Ly\alpha}_{\rm optical}$ $\times$ $\frac{6564.61}{1215.67}$. 
We place both blobs in the same velocity frame using the redshifts
obtained from H$\alpha$. Note that all the optical and NIR wavelengths
are converted to vacuum wavelengths and corrected to the heliocentric
frame.
In both blobs, the \lya\ line is {broader} than the H$\alpha$ line and
has a more complicated structure (i.e., an asymmetric profile or multiple
peaks), implying that the \lya\ lines do experience resonant scattering.

LAB01 has a \lya\ line width of $\Delta v_{\rm FWHM}$ =
520\,$\pm$\,40\,\kms\ whose peak coincides with that of the H$\alpha$
line. While the blue side of the \lya\ profile ($v < 0$\,\kms) agrees
well with that of the H$\alpha$ line, it has an extended red wing up to
at least $v$ $\sim$ +600\,\kms.  It is not clear whether this red wing
arises from the several poorly-resolved red peaks.
%
% suggested by the arrows in Figure \ref{fig:line_comparison}.
%
LAB02 shows a double peaked \lya\ profile with a stronger red peak,
which is often observed in high-$z$ \lya\ galaxies \citep{Tapken07}.
LAB02's \lya\ profile is broad, extending from $\sim$$-800$\,\kms\ to
$\sim$$+$1000\,\kms.  The stronger red peak itself has a line width of
$\Delta v_{\rm FWHM}$ $\sim$ 420\,\kms.  LAB02's \lya\ profile also has
a sharp absorption feature (or lack of emission) at $v \sim 0$\,\kms\
that coincides with its H$\alpha$ line center.

To quantitatively measure the velocity offset between the \lya\ and
H$\alpha$ lines, we must first determine each line's center.
For the asymmetric \lya\ line, we measure the wavelength at the peak flux,
which is consistent with the methodology of \citet{Steidel04, Steidel10}.
Because the \lya\ profile is noisy, we smooth the spectrum with a boxcar
filter of 3 spectral pixels ($\Delta v$ $\simeq$ 40\,\kms) and measure
the wavelength of the brightest flux.
For the symmetric H$\alpha$ profiles in both blobs, we fit the spectrum
with a Gaussian profile including the neighboring \NII\ lines.
We fit three Gaussian components with the same velocity width, but
different intensities, centered at 6563\AA\ (H$\alpha$), 6549\AA, and
6583\AA\ ([\ion{N}{2}]).  
As mentioned earlier, due to the slight contamination of LAB01A's spectrum
with light from LAB01B, we simultaneously fit the observed H$\alpha$
profile for both galaxies and show only the spectrum of galaxy A in
Figure \ref{fig:line_comparison}.

%----------------------------------------------------------------------
\begin{deluxetable*}{ccccccc} % emulateapj
%\setlength{\belowcaptionskip}{600pt}
%\begin{deluxetable}{ccccccc}
%\tablewidth{0pt}
\tabletypesize{\small}
\tabletypesize{\scriptsize}
%\rotate
\tablecaption{Properties of H$\alpha$ Lines}
\tablehead{
%--------------------
\colhead{Target}&
\colhead{}&
\colhead{$z$}&
\colhead{$L$(H$\alpha$)}&
\colhead{$\sigma$(H$\alpha$)}&
%\colhead{Log(\bptx)}&
\colhead{\dvlya}\\
%--------------------
\colhead{}&
\colhead{}&
\colhead{}&
\colhead{($\times10^{43}$\unitcgslum)}&
\colhead{(\kms)}&
%\colhead{}&
\colhead{(\kms)}
}
%--------------------
\startdata
% Target     &    &       $z$                &   $L$(H$\alpha$)           &  $\sigma$(H$\alpha$) & Log([\ion{N}{2}]/H$\alpha$)        
%            &    &                          & $\times10^{43}$\unitcgslum &                      &        
  CDFS-LAB01 &  A & 2.31421 $\pm$  0.0001    &       1.16  $\pm$ 0.06     & 184 $\pm$  9&\phn\phn$+$0   $\pm$ 20  \\
             &  B & 2.30928 $\pm$  0.0001    &       0.78  $\pm$ 0.05     & 134 $\pm$  8& ...                     \\
  CDFS-LAB02 &  A & 2.31928 $\pm$  0.0001    &       2.42  $\pm$ 0.11     & 152 $\pm$  5&        $+$230 $\pm$ 30  

\enddata
\label{tab:line_properties}
%\tablecomments{}
\end{deluxetable*} % for emulateapj
%----------------------------------------------------------------------

The velocity offset, \dvlya\, is defined as the difference between
the H$\alpha$ line center and the \lya\ peak wavelength.  We find
a negligible offset between \lya\ and H$\alpha$ (\dvlya\ $\simeq$
$+0$\,$\pm$\,20\,\kms) for LAB01A, and an offset, \dvlya\ $\simeq$
$+230$\,$\pm$\,30\,\kms, for LAB02A.
These \dvlya\ are smaller than those of LBGs, which range from
+150\,\kms\ up to +900\,\kms\ with an average of +450\,\kms\
\citep{Steidel10}. Among this LBG sample, only 12\% of galaxies have
\dvlya\ smaller than 225\,\kms, thus the probability of finding \dvlya\
$\lesssim$ 230\,\kms\ for two \lya\ blobs by chance is rather small
($\sim$1.5\%) if our \lya--H$\alpha$ offsets are drawn from the same
\dvlya\ distribution\footnote{We note that it is not clear whether the
distribution of \dvlya\ in \citet{Steidel10} is representative of all
LBGs given its sharp break around 300\,\kms.}.

%----------------------------------------------------------------------
%\input{./img/plot_spec1d_abs.tex}
%\input{f06.tex}
%----------------------------------------------------------------------

%----------------------------------------------------------------------
\begin{figure*}
\epsscale{1.00}
\epsscale{0.90}
%\plotone{plot_spec1d_abs.ps}
\plotone{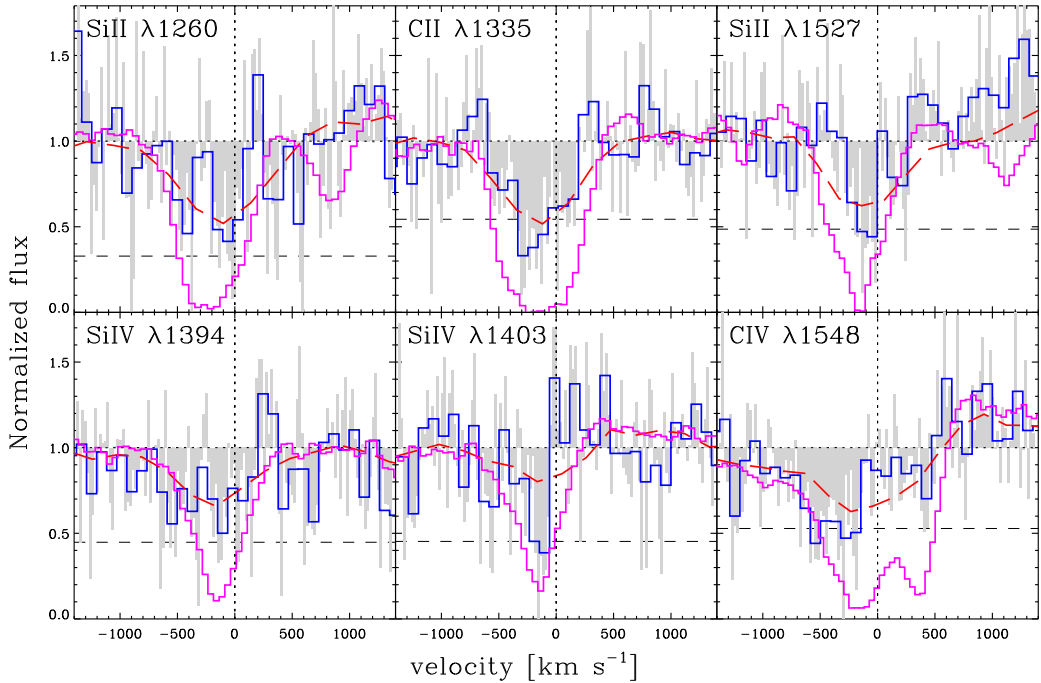}
\caption{
Velocity profiles of absorption lines in LAB02.
(Top) Low-ionization lines : \ion{Si}{2} $\lambda$1260, \ion{C}{2}
$\lambda$1334, and \ion{Si}{2} $\lambda$1526.
(Bottom) High-ionization lines : \ion{Si}{4} $\lambda\lambda$1393,
1402 and \ion{C}{4} $\lambda$1548.
The gray and blue histograms are the unbinned and re-binned (by 4
pixels) histograms, respectively. Red and magenta lines represent the
composite LBG spectrum \citep{Shapley03} and the MS1512-cB58 spectrum
\citep{Pettini00}, respectively.
The dashed horizontal lines indicate the 3$\times${\it rms} noise of
the continuum.  We use the low-ionization interstellar (IS) lines to
measure velocity offsets, \dvis\ $\sim$ $-$100\,\kms\ to $-$200\,\kms,
which are consistent with the modest outflow velocity inferred from the
\lya\ and H$\alpha$ line comparison.
}
\label{fig:absorption}
\end{figure*}
%----------------------------------------------------------------------

%----------------------------------------------------------------------

For our two blobs, \dvlya\ is either consistent with zero (LAB01) or
somewhat redshifted (LAB02).    Therefore, according to the simplest
picture, we find no evidence for a blueshift and thus for gas infall in
either blob.
The absence of a redshifted \dvlya\ in LAB01 does not necessarily exclude
an outflow, because LAB01A's \lya\ profile has an extended red wing
and/or multiple red peaks, which can be outflow signatures.
If we interpret the observed \dvlya\ as a proxy for outflow velocity,
then, at face value, any outflows in LAB01 and LAB02 are weaker than
those typical of LBGs.  In \S\ref{sec:discussion}, we further consider the
implications of our \dvlya\ results using radiative transfer calculations.

%----------------------------------------------------------------------
\subsection{Interstellar Absorption Lines}
\label{sec:absorption}

In addition to the \lya\ emission line, the optical spectrum of LAB02
shows several absorption lines (Figure \ref{fig:spec1d}).  These
absorption lines provide an opportunity to constrain the kinematics of
cold gas on the near side of the galaxy, i.e., between the galaxy and
observer, and thus to verify the detection of the modest \dvlya\ outflow
signature in the previous section.  Note that the systemic velocity from
H$\alpha$ spectroscopy is again critical.

Because the S/N of the continuum in the LAB02 spectrum is very low,
we first re-bin the spectrum by four spectral pixels, which worsens the
resolution of the final velocity bins to $\sim$90 \kms.
Figure \ref{fig:absorption} shows the velocity profiles of six
absorption lines whose peaks were detected at the $\gtrsim$ 3$\sigma$
level in the rebinned spectrum.  We show the absorption profiles of
three low-ionization interstellar lines (\ion{Si}{2} $\lambda$1260.42,
\ion{C}{2} $\lambda$1334.53, and \ion{Si}{2} $\lambda$1526.72) and three
high-ionization lines (\ion{Si}{4} $\lambda\lambda$1393.76, 1402.77 and
\ion{C}{4} $\lambda$1548.20). For comparison, we also plot the composite
LBG spectrum ($R\sim800$) from \citet{Shapley03} and the individual
spectrum of MS1512-cB58 ($R \sim 1300$) which is one of the apparently
brightest LBG whose light is amplified due to the gravitational lensing
\citep{Pettini00}.
Although the S/N is low, the velocity ranges spanned by LAB02's absorption
lines roughly agree with these two templates.

We find that all interstellar absorption lines are blueshifted against
the systemic velocity, indicating that the intervening gas is outflowing,
which is consistent with the observed \dvlya\ outflow signature discussed
in the last section.  To estimate the velocity of the outflowing material,
we measure how much the centroids of these velocity profiles are offset
from the H$\alpha$ line: \dvis.
While the high-ionization lines (\ion{Si}{4} $\lambda\lambda$1393, 1402
and \ion{C}{4} $\lambda$1548) have the highest S/N, they are known to be
contaminated with broader absorption features arising from stellar winds
from massive stars \citep{Leitherer95,Shapley03}.  Therefore, we measure
the centroid velocities (\dvis) using the three low-ionization lines.
Two low-ionization lines (\ion{C}{2} $\lambda$1334 and \ion{Si}{2}
$\lambda$1526) have \dvis\ $\sim$ $-$100\,\kms\ to $-$200\,\kms.
(It is difficult to reliably measure the centroid of the \ion{Si}{2}
$\lambda$1260 line.)
The absorption profile of \ion{C}{2} $\lambda$1334 might extend up to
$|v_{\rm max}|$ $\sim$ 600\,\kms\ blueward, although its low S/N precludes
a reliable measurement.  If we adopt its centroid velocity as
an outflow velocity, then it is similar to that of LBGs, which
average $\langle v_{\rm IS} \rangle$ = $-164$\,\kms, but can be as large
as $-$500\,\kms\ \citep{Steidel10}.

%----------------------------------------------------------------------
\subsection{Velocity Dispersions of Embedded Galaxies}
\label{sec:dynamical_mass}

By fitting Gaussian profiles to the integrated H$\alpha$ profiles (Figure
\ref{fig:linefit}), we obtain the velocity dispersions, $\sigma_{\rm
H\alpha}$ = 184\,$\pm$\,9\,\kms\ and 134\,$\pm$\,8\,\kms\ for the
LAB01A and LAB01B, respectively. For LAB02A, $\sigma_{\rm H\alpha}$ =
152\,$\pm$\,5\,\kms. These velocity dispersions are corrected for the
instrumental resolution ($\sigma_{\rm res}$ $\sim$ 32\,\kms).

The internal velocity dispersions of the embedded galaxies ($\sigma_v$
= 130 -- 190\,\kms) are high compared to those of star-forming galaxies
at $z\sim2$. \citet{Erb06b} find that LBGs at $z\sim2$ have an average
of $\langle\sigma\rangle$ = 108\,\kms\ with a standard deviation
of 41\,\kms, excluding galaxies with AGNs.  Our lowest $\sigma_{\rm
H\alpha} \sim 130$\,\kms\ corresponds to the upper $\sim$30\% percentile
of their distribution.  In particular, there is only one galaxy in
the \citet{Erb06b} sample that has higher $\sigma$ than LAB01A does
($\sigma_{\rm H\alpha}$ = 190\,\kms).

%----------------------------------------------------------------------
\section{Discussion}
\label{sec:discussion}

In this section, we compare the observed \lya\ profiles and their
offsets from the H$\alpha$ lines with the predictions of \lya\
radiative transfer (RT) calculations \cite[][hereafter D06, V06 and V08,
respectively]{Dijkstra06a, Verhamme06, Verhamme08}.
To test the scenario in which gas in the surrounding cloud cools and
flows toward the blob center, we compare our data with the spherical
collapse model in D06.  For the picture in which gas outflows from the
embedded galaxies in the blob's core, we employ the expanding shell
model investigated by V06.
Note that these RT calculations assume that infall and outflow occur in
spherical symmetry, in a collapsing sphere for the inflow model and in
an expanding shell for the outflow model.  As we argued previously, even
such simple models should reproduce the general behavior of the \lya\
profile as long as the basic assumptions listed in  \S\ref{sec:shift}
are valid.
In \S\ref{sec:future}, we examine those assumptions in detail and discuss
the future work required to overcome these limitations.

\subsection{Absence of Blueshift:  No Gas Infall?}
\label{sec:infall}

In \S\ref{sec:shift}, we found that the \lya\ lines of the two blobs
are not blueshifted relative to the H$\alpha$ line centers within the
measurement uncertainties.  The absence of a blueshift rules out the
simplest infall model where a spherical gas cloud is collapsing onto a
central \lya\ source or where the collapsing cloud itself is emitting
\lya\ photons.
In this model, \citet{Verhamme06} and \citet{Dijkstra06a} study \lya\
radiative transfer, which depends on various parameters such as the
location of the \lya\ source, the infall velocity field, the column
density of neutral hydrogen (\NHI), and the Doppler parameter $b$
representing the thermal and turbulent motion of the gas (see Fig.~4 and
7 in D06 or Fig.~5--7 in V06).  
The key prediction of these RT calculations is that the \lya\ spectrum
is double peaked with an enhanced blue peak, producing an effective
blueshift of the \lya\ profile. While the blue peak might be suppressed
by the intervening IGM after the \lya\ photons escape the collapsing
cloud \citep{Dijkstra06b}, this effect is unlikely to be significant here
because the mean transmission blueward of \lya\ should be relatively high
at $z\sim2.3$: $\langle e^{-\tau}\rangle$ $\sim$ 0.8 \citep{McDonald01},
where $\tau$ represents \lya\ optical depth.

The blueshift of \lya\ photons against the systemic velocity should
arise as long as 1) there are infall motions along the LOS and 2) \lya\
photons on their way out of the system encounter optically thick gas with
a velocity opposite to their propagation direction \citep{Dijkstra06a}.
One possible way to hide a blueshift in this simple infall model would
be if the infalling gas is made optically thin by an ionizing source or
has low enough column density so that the \lya\ photons do not suffer
resonant scattering (see Fig.~11 in D06 or Fig.~6 in V06).  In this case,
we should still see nearly symmetric \lya\ profiles or extended blue
wings, neither of which is observed in our \lya\ profiles. Therefore,
we rule out this simple gas infall scenario for our two \lya\ blobs.

In contrast to this simple model, the actual situation is likely to be
more complex and the bulk motion of gas may not be spherically symmetric.
For example, the H$\alpha$ detections suggest that the
galaxies embedded in the blobs are forming stars, which could generate
mechanical feedback into the surrounding gas cloud.  There is also ample
evidence for galactic scale outflows in star-forming galaxies at $z$ =
2--3 \cite[e.g.,][]{Steidel10}.
Thus, one could imagine that even if gas infall takes place over larger
radii (up to $\sim$50 kpc), the innermost part of the gas cloud, close
to the galaxies, could be strongly affected by outflows similar to those
observed in other star-forming galaxies at $z=2-3$.  
In this case, the emerging \lya\ profile will be more sensitive to the
core of the blob, presumably the densest part of the IGM, than to the
gas infall. As a result, it might be difficult to detect the infalling
gas by measuring the \lya\ line shift.
More simulations are required to test whether generic galactic scale
outflows could affect gas infall and whether infall and outflows can
coexist.  This kind of more realistic model has not been considered yet
in RT calculations, so its spectral signatures (if any) are unknown.

The other caveat is that the inflow might occur along filamentary
streams \citep{Keres05,Dekel09} and their covering factor could be small
\citep{Faucher-Giguere&Keres10} such that we might completely miss the
infall signatures in a particular LOS.  We will further discuss this
caveat in \S\ref{sec:misalignment}.

\subsection{Small \dvlya: Weak or No Outflow?}
\label{sec:outflow}

In \S\ref{sec:shift}, we found that the offset of the \lya\ line relative
to H$\alpha$ was either consistent with zero (LAB01) or redshifted
(LAB02).  The existence of a modest outflow in LAB02 was further
supported by the low-ionization absorption line shifts discussed in
Section \ref{sec:absorption}.  The absence of a redshifted \dvlya\ in
LAB01 does not necessarily exclude an outflow because of its extended
\lya\ profile toward the red.  Here we compare the \lya\ profiles of
our blobs with the RT outflow models of \citet{Verhamme06, Verhamme08}
to interpret the velocity offset \dvlya\ in a more sophisticated manner.

In these radiative transfer calculations, the \lya--H$\alpha$ velocity
offset is modulated entirely by the radiative transfer of \lya\ photons
through a continuous medium within the shell.  Another possibility,
which we discuss in Section \ref{sec:clumpyCGM}, is that the gas around
the galaxy is clumpy and that the bulk motions of these clumps primarily
modulate the emerging \lya\ profile \citep{Steidel10}.
In the following section, we constrain the outflow velocity or ``wind
velocity'' from the embedded galaxies using the RT calculations.  It is
then possible to ask whether such an outflow is energetic enough (i.e.,
the expansion velocity \vexp\ is large) to expel/launch gas out to larger
radii, thus creating the extended \lya\ emission observed as a blob with
a typical size of $\gtrsim$ 50kpc.

\subsubsection{Comparison with Radiative Transfer Models}

To connect \dvlya\ with an outflow velocity, we compare each blob's  \lya\
profile with a model with an expanding shell geometry.  Such a geometry
would arise in the simple ``superwind'' or ``hyperwind'' model for \lya\
blobs proposed by \citet{Taniguchi&Shioya00} and \citet{Taniguchi01}.
In this scenario, supernova explosions or stellar winds following an
intense starburst in galaxies develop into a so-called superbubble. If
the kinetic energy deposited into the surrounding gas overcomes the
gravitational potential energy of the galaxy, the gas clouds are blown
out into intergalactic space as a superwind (e.g., Heckman et al. 1990).
The gas blown out from the galaxies forms a shell, and the \lya\
photons are scattered from this expanding shell (with \vexp), escaping
preferentially by the scattering from the backside.  In contrast, the
H$\alpha$ lines provide the systemic velocity of the star forming region.

V06 and V08 demonstrate that the radiative transfer of \lya\ in this
simple shell geometry is able to explain a wide range of observed \lya\
profiles.  V08 classified \lya\ profiles into three categories:
(1) single-peaked asymmetric profile with an extended red wing,
(2) profile with double peaks separated by sharp absorption 
    at $v \sim 0$\,\kms, and
(3) asymmetric profile with bluer bump(s).
Thanks to the H$\alpha$ line, which provides the velocity center, we
can determine that LAB01 and LAB02 correspond to (1) and (2) without
ambiguity.

We start with LAB01, which has an asymmetric \lya\ profile without
any obvious blue peaks, but with an extended red wing and/or multiple
red peaks (i.e., category 1 above).  V06 show that a blue peak ($v$
$<$ 0\,\kms) is strongly suppressed with increasing ($v_{\rm exp}/b$)
ratio, i.e., when the bulk outflow motion becomes more important than the
thermal and turbulent motion of the shell gas.  The resonant scattering
in the expanding shell produces mainly two red peaks depending on the
paths  of the emergent photons:  one peak near the line center ($v$
$\sim$ 0\,\kms) and another at $v \simeq 2 v_{\rm exp}$, corresponding
to photons that experience zero or one back-scattering from the receding
shell, respectively.  Here, the back-scattering of a photon means that
it travels across the empty interior before re-entering the shell at
a different location.  Thus the location of the second red peak is a
useful diagnostic for the outflow velocity.

Because of the degeneracy between many model parameters (\vexp, $b$,
\NHI, $\tau_{\rm dust}$), constraining the outflow velocity well requires
fitting the spectra directly to the RT calculations.  We do not have
the means to do that here, so we estimate \vexp\ for LAB01 from how much
its red \lya\ peaks are redshifted against its H$\alpha$ line center.
If \NHI\ is sufficiently large (\NHI $\gtrsim$ $10^{20}$\,cm$^{-1}$),
the second red peak becomes larger and merges with the first red peak,
so that the stronger red peak measures 2\,\vexp\ (see Fig.~1 of V08).
If this case applies here, our \dvlya $\sim$ 0\,\kms\ implies \vexp $\sim$
0\,\kms, i.e., that the shell is static.

On the other hand, decreasing \NHI, increasing dust extinction ($\tau_{\rm
dust}$), and decreasing $v_{\rm exp}/b$ tends to strengthen the first red
peak, so that the brighter peak is located at 0 $\lesssim v\lesssim$
\vexp\ and does not carry any \vexp\ information.  In this case,
the location of the weaker second peak provides an estimate of \vexp.
As discussed in \S\ref{sec:confirmation}, the LAB01 profile's extended
red wing could be interpreted as multiple components (see Figure
\ref{fig:line_comparison}).  If we adopt the small bump at $v$ $\sim$
300\,\kms\ as the second peak, the expansion velocity of the shell is
\vexp\ $\sim$ 150\,\kms.  Considering both cases, the outflow velocity
in LAB01 could range from 0\,\kms\ to $\sim$\,150\,\kms.

The \lya\ profile of LAB02 is most consistent with the double-peaked
profile (i.e, category 2 above) of \citet{Verhamme08}, who claim that such
profiles arise when the shell is almost static (\vexp\ $\sim$ 0\,\kms).
Due to the H$\alpha$ measurement, there is no ambiguity about the location
of the velocity center. Even if we consider only the red side of the
profile and classify it as a single asymmetric profile like LAB01,
we would obtain an outflow velocity only as high as $\sim$ \dvlya/2 =
115\,\kms.  The blueshifted centroids of the low-ionization absorption
lines in LAB02 are also consistent with this inferred outflow velocity.

Therefore, if the expanding shell geometry is applicable to our two
\lya\ blobs, we conclude that any outflows from the embedded galaxies
are not strong (\vexp\ $\lesssim$ 150\,\kms) compared to the values
\cite[$\sim$ 500--1000\,\kms; see Eq. (2) of][]{Taniguchi&Shioya00}
required by the extreme ``superwind'' or ``hyperwind'' scenarios
\citep{Taniguchi&Shioya00, Taniguchi01, Ohyama03} to explain the large
size ($r \gtrsim 50$\,kpc) and broad \lya\ linewidth (FWHM $\sim$
1000\,\kms) of blobs.

The simple shell geometry assumed here has some physical motivation
and has been explored by extensive RT calculations (e.g., V06 and V08).
However, this model implies that the outflows are intermittent with a
small duty cycle, contrary to the observation that outflows are seen in
most star forming galaxies at high redshift.  We note that if there is
a continuous stream of outflowing material, it might lead to a different
interpretation of kinematic signatures discussed above.

One other claim of an outflow in a blob was reported by \citet{Wilman05}
using IFU observations of the \lya\ line in Steidel's Blob 2.  Using
an expanding shell model, they found that the central wavelength,
i.e., velocity, of absorption systems on top of the underlying broad
\lya\ profiles is coherent across the whole blob with a blueshift
of $\sim$250 \kms.  Because the systemic velocity of the blob gas is
unknown in their case, there are other possible interpretations of their
data, even including an inflow \citep{Dijkstra06b}.  \citet{Bower04}
and \citet{Weijmans09} found that Steidel's Blob 1 has an extremely
complicated velocity structure, including velocity shears around the
embedded galaxies that can be interpreted as either outflow or rotation.
It is interesting that all investigations of blob kinematics to date
(in our CDFS-LAB01 and CDFS-LAB02, Steidel's Blobs 1 and 2) suggest
relatively small outflow velocities.  Therefore, the large wind speeds
required by super/hyper-wind models to explain the extended \lya\
emission are excluded.  One way to reconcile the small outflows with the
``superwind'' model is to assume that the outflow was once more energetic
and that the expanding shell has been broken up or slowed by contact
with the surrounding IGM \citep{Wilman05, Weijmans09}.
Clearly, constraining the kinematics of more \lya\ blobs is needed to
address whether these small outflows are representative.

\subsubsection{Summary}

We conclude that the gas motion around the galaxies embedded in these
\lya\ blobs is not consistent with models like the simple superwind
picture \citep{Taniguchi&Shioya00} that assume a spherical shell geometry
and a shell velocity exceeding $\sim$500 \kms.
Yet galactic scale outflows from star-forming galaxies at $z=2-3$ appear
to be common.  The outflow properties have been extensively studied by
comparing the \lya\ velocity center with those of optically-thin nebular
lines and/or interstellar absorption lines \cite[e.g.,][]{Pettini01,
Shapley03, Steidel04, Steidel10}.
Recently, \citet{Steidel10} show that \lya\ lines from LBGs are redshifted
against H$\alpha$ lines up to several hundreds \kms\ with an average
of $\avg{\Delta v_{\rm Ly\alpha}}$ = +445\,\kms, while the interstellar
absorption lines are blueshifted against the nebular lines by $\avg{\Delta
v_{\rm IS}}$ = $-164$\,\kms.

While these \dvlya\ and \dvis\ offsets are interpreted as an outflow
signature, it is not clear how the absorbing material is distributed
around galaxies.
{\it Were the gas blown out in a spherical shell}, as in the superwind
picture discussed above \cite[e.g.,][V08]{Taniguchi&Shioya00,
Schaerer&Verhamme08}, we could directly compare the observed \dvlya\
of blobs and LBGs as a proxy for \vexp.  The velocity offsets of our two
\lya\ blobs, \dvlya\ $\sim$ 0\,\kms\ and 230\,\kms, are small in relation
to LBGs \citep{Steidel10}; of 41 LBGs at $z$ = 2 -- 2.6 with both \lya\
and H$\alpha$ spectroscopy, $\sim$88\% have \dvlya\ $>$ 225\,\kms.
Therefore, any wind arising from galactic star formation has a lower
expansion velocity in {these two LABs} than is typical of LBGs.

\bigskip
\subsection{Caveats}
\label{sec:future}

We have adopted two basic assumptions in our analysis:  that there is
inflowing or outflowing gas in the LOS to the blob and that the H$\alpha$
line marks the kinematic center of the blob gas.  Here we examine how
these assumptions may affect our conclusions and comment on the recent
model suggested by \citet{Steidel10} to explain the circum-galactic
kinematics of LBGs.

\medskip
\subsubsection{Misalignment of Gas Flow to Line of Sight}
\label{sec:misalignment}

We have assumed that there is gas inflowing or outflowing along the LOS
to the embedded galaxies that lie in our spectroscopic slit.  In this
case, the observed \lya\ photons resonantly scatter in this optically
thick infalling/outflowing material until escaping into the LOS of
the observer.  However,  we do not know how the blob gas is distributed
around the embedded galaxies nor whether it is clumpy or smooth.

For example, the gas flow may not be isotropic.  As in bipolar outflows
\cite[e.g.,][]{Bland&Tully88}, a galactic-scale outflow may occur in the
direction of minimum ISM pressure, which is typically perpendicular to
the stellar disk.
Or, the covering factor of the outflow material, the fraction of emission
intercepted by the absorbers, could be rather high, at least 0.5 or up to
unity as observed in nearby AGNs \citep{Crenshaw03}.
If gas accretion is taking place instead \citep{Keres05,
Keres09}, numerical simulations suggest that the infall may occur
preferentially along filamentary streams \citep{Dekel09, Goerdt10}.

If the bulk motion of gas (either infalling or outflowing) happens
to be mis-aligned with our LOS, we may underestimate or even fail
to detect the relative velocity shifts.  Therefore, it is critical to
expand the \lya--H$\alpha$ comparison to a larger sample to average over
any geometric effects and obtain better constraints on the incidence,
direction, speed, and isotropy of bulk gas motions in blobs.  If no
large values of \vexp\ are measured, even for a larger sample, outflows
are unlikely to push the blob gas to larger radii, where it can be
illuminated.  On the other hand, an absence of blueshifted \lya\ lines
would argue that infall is not dominant, at least not near the galaxies.

Recent cosmological simulations including radiative transfer \cite[e.g.,]
[]{Faucher-Giguere10} are now able to examine the statistics of blobs
whose extended \lya\ emission arises from cooling radiation. These
models can predict the range of gas accretion covering factors, surface
brightness profiles and ratio of blue-to-redshifted \lya\ lines for blobs.
A larger sample of blobs with measured \lya\ and H$\alpha$ kinematics
would provide a new test of these blob emission models.

\bigskip
\subsubsection{Misidentification of Kinematic Center}

Our second key assumption is that the H$\alpha$ line marks the kinematic
center of the surrounding gas, i.e., the terminus of any infall or the
origin of any outflow.
For an outflow, this assumption is reasonable because all possible
causes of the outflow (e.g., stellar winds, SNe, or AGNs) lie in or
about galaxies, as does the H$\alpha$ line itself.  For similar reasons,
this assumption is likely to be valid for gas accretion (infall) into
an embedded galaxy.  However, it is possible that the accreting gas
responds to the overall DM potential rather than to individual galaxies.
Thus one could expect that detecting gas infall is more difficult than
finding outflows using the technique presented in this paper.

To overcome this issue, it is critical to increase the sample size and
measure how each blob's \lya\ profile varies spatially.  This experiment
should be conducted along multiple sightlines using using different
H$\alpha$ sources found in the same blob.
In principle, it is possible to compare the offsets between the extended
\lya\ and and the {\it extended} H$\alpha$ lines in two dimensions,
i.e., between the H$\alpha$ and \lya\ screens.   Unfortunately,
the H$\alpha$ surface brightness is expected to be $\sim$3\% -- 10\%
of \lya\ depending on the energy source.  Thus, it will be challenging
to obtain 2D H$\alpha$ IFU measurements in the near future.

Currently, our \lya--H$\alpha$ comparison includes only a handful of \lya\
blobs because of the limited redshift range where NIR spectroscopy of
the optically-thin nebular lines (e.g., H$\alpha$ and the [\ion{O}{3}]
$\lambda$5007 line) is allowed.  However, future CO observations with
ALMA will also provide the systemic velocity of the embedded galaxies,
permitting this test to be expanded to a larger sample of \lya\ blobs.

\subsubsection{Relationship to Clumpy, Circum-Galactic Medium of LBGs?}
\label{sec:clumpyCGM}

As we were writing this paper, \citet{Steidel10, Steidel11}
proposed a model for LBGs in which cool gas is distributed around the
galaxy and accelerates radially outward with increasing galactocentric
distance.
They suggest that this blown-out material extends to $\sim$80\,kpc,
comparable to the typical size of \lya\ blobs, and that all LBGs would
be classified as blobs if narrowband images were sensitive to surface
brightness thresholds 10 times lower than typical of blob/emitter surveys
(i.e., $\sim$$10^{-19~}$\unitcgssb).
Is this ``clumpy circum-galactic medium (CGM) model'' consistent
with the properties of our \lya\ blobs?  Unlike in the RT models that
we considered earlier, the gas in this model is composed of small clumps
\cite[e.g.,][]{Neufeld91,Hansen&Oh06}.

A critical part of the \citet{Steidel10} analysis involves stacking
spectra of apparent galaxy-galaxy pairs at $z\sim2-3$ to obtain a high
S/N interstellar absorption profile arising from the gas surrounding
the foreground galaxies. Sightlines to the background galaxies provide
information on the spatial distribution of circum-galactic gas as a
function of the galactocentric distances.
There are no background galaxies behind our two blobs, and even
constraining the absorption profile in LAB02A was challenging.  We do
not detect a stellar continuum in LAB01A or LAB01B. As a result, we are
unable to test whether the \citet{Steidel10} LBG model could explain
the kinematics of either of our \lya\ blobs.  Nevertheless, we add a
few qualitative comments here.

In the \citet{Steidel10} model, \lya\ photons from the central galaxy are
scattered off the {\it surface} of these discrete clumps; thus photons
acquire a Doppler shift corresponding to the velocity of each clump.
\lya\ photons can escape the system when they achieve velocity offsets
large enough to take the photons off-resonance for any material between
the last scattering clump and the observer.
Therefore, the \lya\ line shift is due to bulk motions of the clumpy
material rather than to radiative transfer effects where \lya\ photons
experience numerous resonant scatterings before escaping the cloud
(Section \ref{sec:outflow}).
The profiles of \lya\ and the low-ionization interstellar absorption
lines are determined by the velocity field and covering fraction of these
gas clumps. The apparent peak of \lya\ emission is modulated primarily
by the clumps with $v \sim 0$: \lya\ emission is more redshifted and
weaker for a wider velocity range spanned by the interstellar absorption.
The velocity of the clumpy medium continually increases outward, unlike
in the radiative transfer models discussed earlier, where the outflowing
shell had one velocity.
In the \citet{Steidel10} model, the velocity range spanned by the
absorption lines reflects that of the surrounding gas distribution, and
the \lya--H$\alpha$ offset does {\it not} represent the outflow velocity.

If this {clumpy CGM} model is applicable to blobs, then our
observations of a small \lya--H$\alpha$ offset could indicate an absence
of neutral gas at $v \sim 0$ rather than the weak gas outflow suggested
by the RT models.
One test would be to constrain the interstellar absorption line profile,
which should extend to larger blueward velocities, but lack a component
around $v\sim0$\,\kms, if the clumpy model is accurate.
Clearly, higher S/N spectra to reliably measure these weak absorption
lines are required.

The most notable outcome of the large spatial extent of a clumpy CGM
around star-forming galaxies is diffuse \lya-emitting halos.  Extended
\lya\ emission has been detected in stacked narrowband images of LBGs
by \citet{Hayashino04}.
More recently, by stacking deeper narrowband \lya\ images of LBGs,
\citet{Steidel11} show that all classes of star-forming galaxies
exhibit diffuse \lya\ halos out to projected radii of $\sim$80\,kpc,
regardless of their spectroscopic properties (i.e., either absorption
or emission in \lya\ within a spectroscopic slit), although these halos
are 10$\times$ lower surface brightness than typical blobs.  The \lya\
surface brightness profiles of the stacked images are similar among LBG
subclasses, with only different intensity scalings.
Thus, \lya\ blobs may represent the most extreme example of galaxies
with diffuse \lya\ halos.
The key difference of the clumpy CGM model is that the diffuse \lya\ halo
originates from \lya\ photons scattered from the galaxy's \ion{H}{2}
region(s) rather than from external mechanisms such as photoionization
by AGN, photoionzation by metagalactic UV background radiation, or via
cooling radiation.

At this time, the applicability of the clumpy model to \lya\ blobs is
unclear.  Any comprehensive model must explain the differences between
our blobs and LBGs:
1) a comparison of their number density and clustering properties suggests
that {blobs could occupy higher density environments} \cite[e.g.,][]{Yang10},
2) {the extended emission of blobs is typically 10 times brighter,}
3) the \lya--H$\alpha$ offset of our blobs is lower.
The first and second differences could arise if the blobs host
more massive galaxies and higher star formation rates than typical of
star-forming galaxies at high-$z$.  It remains to be tested whether
star formation in the galaxies embedded in blobs can produce enough
\lya\ photons to explain the higher luminosity and surface brightness
of \lya\ halos.
The \lya--H$\alpha$ offset is clearly tied in some way to the extended
\lya\ emission of our blobs, as it is unusually low for both blobs
compared to the offset distribution for LBGs \cite[see also][for compact
\lya\ emitters]{McLinden10}.
It would interesting to check whether the LBG subclasses in \citet{Steidel11}
 with {brighter} extended emission have smaller \lya--H$\alpha$ offsets
and weaker IS absorption at $v\sim0$\,\kms\ than the others.
For now, we note that the RT outflow model discussed in Section
\ref{sec:discussion} adequately explains the sense and magnitude of
both the \lya--H$\alpha$ and the H$\alpha$-interstellar absorption line
offsets in LAB02, whereas the \citet{Steidel10} model would require this
agreement of the offsets to be a coincidence.
Of course, we are only able to perform this test using one blob at present.
As stressed earlier, a larger sample of \lya--H$\alpha$ offsets is
required to test whether smaller \lya--H$\alpha$ offsets are generic
properties of \lya\ blobs.

%\clearpage
%----------------------------------------------------------------------
\section{Conclusions}
\label{sec:conclusion}

Determining the source of \lya\ emission in extended nebulae at
high redshift requires that we measure their gas kinematics with an
optically-thin line such as \halamb, which, unlike the optically-thick
\lya\ emission, is not much altered by radiative transfer effects and
is more concentrated about the blob's core.  The comparison of the \lya\
and H$\alpha$ line centers then constrains any infall or outflow of the
\lya-emitting gas relative to the systemic velocity of the blob defined
by H$\alpha$.  We obtain optical and NIR spectra of the two brightest
\lya\ blobs (CDFS-LAB01 and CDFS-LAB02) from the \citet{Yang10} sample
using the Magellan/MagE optical and VLT/SINFONI NIR spectrographs.
Both the \lya\ and H$\alpha$ lines confirm that these blobs lie at the
survey redshift, $z \sim 2.3$ ($\lambda_c \approx 4030$\AA).

The blobs contain several H$\alpha$ sources, which roughly coincide
with galaxies in {\sl HST} rest-frame UV images.  The H$\alpha$
detections of these embedded galaxies reveal large internal velocity
dispersions ($\sigma_v = 130 - 190$\,\kms).  Furthermore, in the
one system (LAB01) where we can reliably extract profiles for two
different H$\alpha$ sources, their velocity difference is $\Delta v$
$\sim$ 440\,\kms.  The presence of multiple galaxies within the blobs,
and those galaxies' large velocity dispersions and large relative motion,
supports the result of \citet{Yang10} that \lya\ blobs inhabit massive
dark matter halos that will evolve into rich clusters today.  The embedded
galaxies may represent the precursors of brightest cluster galaxies.

To determine whether the gas near the embedded galaxies is predominantly
infalling or outflowing, we compare the \lya\ and H$\alpha$ line centers.
\lya\ is not offset (\dvlya\ = +0\,\kms) in LAB01 and redshifted by
only +230\,\kms\ in LAB02.  These offsets are small compared to those
of Lyman break galaxies, which average +450\,\kms\  and extend to about
+700\,\kms\ \citep{Steidel10}.
By comparing the observed \lya\ profiles with simple radiative transfer
models, we constrain the outflow velocity from the embedded galaxies
to  be $\sim$\,0 -- 150\,\kms\ in the two blobs.  In LAB02, we detect
low-ionization, interstellar absorption lines, including \ion{C}{2}
$\lambda$1334 and \ion{Si}{2} $\lambda$1526, whose blueward shifts of
$\sim 200$\,\kms\ are consistent with the small outflow implied by
the redward shift of \lya.  Models with simple infall geometries or
that require outflow velocities exceeding $\sim$500\,\kms\ \cite[e.g.,
super/hyper-winds discussed in][]{Taniguchi&Shioya00} are ruled out.
Because of the unknown geometry of the gas distribution and the
possibility of multiple sources of \lya\ emission embedded in the blobs,
a larger sample and more sophisticated models are required to test other
infall and outflow scenarios.

\acknowledgments

We thank anonymous referee for his/her helpful comments.
We thank Christy Tremonti, Dusan Kere{\v s} and Mark Dijkstra for useful
discussions. We also thank Max Pettini for providing the spectrum of
MS1512-cB58.
A.I.Z.\ thanks the Max-Planck-Institut f\"ur Astronomie and the Center
for Cosmology and Particle Physics at New York University for their
hospitality and support during her stays there. Y.Y.\ and A.I.Z.\
acknowledge support from the NSF Astronomy and Astrophysics Research
Program through grant AST-0908280 and from the NASA Astrophysics Data
Analysis Program through grant NNX10AD47G.
K.J.\ is supported through the Emmy Noether Programme of the German
Science Foundation (DFG).

\medskip
Facilities: \facility{VLT (SINFONI), Magellan (MagE), Blanco (MOSAIC II)}

%\clearpage

\end{document}